\documentclass[12pt,a4paper,twoside]{article}

\usepackage[dvips]{graphics}
\usepackage{graphicx}
\usepackage{cite}
\usepackage{amsmath}
\usepackage{amssymb}
\usepackage{rotating}

\usepackage [pagewise, mathlines, displaymath]{lineno}

\def\be{\begin{equation}}
\def\ee{\end{equation}}
\def\bea{\begin{eqnarray}}
\def\eea{\end{eqnarray}}

\newcommand{\bbbar}     {\ensuremath{\mathrm{b\bar{b}}}}
\newcommand{\qqbar}     {\ensuremath{\mathrm{q\bar{q}}}}

\newcommand{\epem}              {\ensuremath{\mathrm{e^+e^-}}}

\newcommand{\Nf}      {\ensuremath{n_{\mathrm{f}}}}
\newcommand{\yf}      {\ensuremath{y_{\mathrm{f}}}}
\newcommand{\yc}      {\ensuremath{y_{\mathrm{cut}}}}

\newcommand{\vlarge}  {\phantom{$\Big|$}}

\newcommand{\CA}                {\ensuremath{C_{\mathrm{A}}}}
\newcommand{\CF}                {\ensuremath{C_{\mathrm{F}}}}

\newcommand{\nf}                {\ensuremath{n_{\mathrm{f}}}}
\newcommand{\as}                {\ensuremath{\alpha_\mathrm{S}}}

\newcommand{\asrs}                {\ensuremath{\alpha_\mathrm{S}(\sqrt{s})}}
\newcommand{\ash}               {\ensuremath{\hat{\alpha}_\mathrm{S}}}

\newcommand{\asmu}              {\ensuremath{\alpha_\mathrm{S}(\mu)}}

\newcommand{\asmz}              {\ensuremath{\alpha_\mathrm{S}(\mathrm{M}_{\mathrm{Z^0}})}}

\newcommand{\mz}                {\ensuremath{{\mathrm{M_{Z^0}}}}}

\newcommand{\chisq}     {\ensuremath{\chi^2}}
\newcommand{\chisqd}    {\ensuremath{\chi^2/\mathrm{d.o.f.}}}
\newcommand{\xmu}               {\ensuremath{x_{\mu}}}

\newcommand{\xmuopt}               {\ensuremath{x_{\mu}^\mathrm{opt}}}

\newcommand{\ycut}              {\ensuremath{y_{\mathrm{cut}}}}

\newcommand{\stat}              {\ensuremath{\mathrm{(stat.)}}}

\newcommand{\expt}               {\ensuremath{\mathrm{(exp.)}}}
\newcommand{\had}               {\ensuremath{\mathrm{(had.)}}}
\newcommand{\theo}              {\ensuremath{\mathrm{(theo.)}}}

\newcommand{\yy}                {\ensuremath{y_{23}}}

\newcommand{\rs}                {\ensuremath{\sqrt{s}}}

\newcommand{\invpb}     {\ensuremath{\mathrm{pb}^{-1}}}

\newcommand{\py}                {PYTHIA}
\newcommand{\hw}                {HERWIG}
\newcommand{\ar}                {ARIADNE}

\newcommand{\result} {\ensuremath{\asmz=0.1199\pm0.0010\stat\pm0.0021\expt\pm0.0054\had\pm0.0007\theo}}
\newcommand{\restot} {\ensuremath{\asmz=0.1199\pm0.0060~(\mathrm{total~error})}}

\newcommand{\resultxmuopt} {\ensuremath{\asmz=0.1204\pm0.0009\stat\pm0.0021\expt\pm0.0059\had\pm0.0008\theo}}

\newcommand{\resultthirty} {\ensuremath{\as(34.8\, \mathrm{GeV})=0.1431\pm0.0010\stat\pm0.0028\expt\pm0.0086\had\pm0.0010\theo}}

\newcommand{\resultnlonlla}{\ensuremath{\asmz=0.1175\pm0.0010\stat\pm0.0020\expt\pm0.0052\had\pm0.0020\theo}}

\begin{document}

\begin{titlepage}


\begin{flushright}
\today
\end{flushright}

\bigskip\bigskip\bigskip

\begin{center}\textbf{
\Large Measurement of the strong coupling \boldmath{\as} from the 
three-jet rate in \epem-annihilation using JADE data }
\end{center}

{\Large \par}
\bigskip
\begin{center}
 {\Large J. Schieck$^{a,b}$, S. Bethke$^{c}$, S. Kluth$^{c}$, 
  C. Pahl$^{c}$, Z.~Tr\'{o}cs\'{a}nyi$^{d}$ and the JADE Collaboration\footnote{The members of
    the JADE collaboration are listed in~\cite{JadeAuthors}}} \\
    \bigskip
 \bigskip \bigskip
\end{center}
\par

\bigskip
\begin{abstract}
We present a measurement of the strong coupling  \as\ using
the three-jet rate measured with the Durham algorithm in \epem-annihilation using 
data of  the JADE experiment at 
centre-of-mass energies between 14 and 44~GeV.  Recent  
theoretical improvements provide predictions of the three-jet
rate in \epem-annihilation at next-to-next-to-leading order.
In this paper a measurement of the three-jet rate is used to determine the strong coupling \as\ 
from a comparison to next-to-next-to-leading order predictions matched with next-to-leading 
logarithmic approximations and yields a value for the strong coupling 
\begin{center}
  \result,
\end{center}
consistent with the world average. \\
\bigskip\bigskip

This paper is dedicated to the memory of Joachim Heintze, a founding member
of the JADE collaboration whose essential contributions to JADE and the field are imperishable.\\
\end{abstract}
\bigskip \bigskip\bigskip\bigskip\bigskip
$^{a}$ Ludwig-Maximilians-Universit\"at M\"unchen, Am Coulombwall 1,\\
D-85748 Garching, Germany \newline
$^{b}$ Excellence Cluster Universe, Boltzmannstr. 2, D-85748
Garching, Germany \newline
$^{c}$ Max-Planck-Institut f\"ur Physik, F\"ohringer Ring 6, 
D-80805 M\"unchen, Germany \newline
$^{d}$ Institute of Physics, University of Debrecen, H-4010 Debrecen P.O.Box 105, Hungary \newline

\bigskip
\bigskip
\vfill
\end{titlepage}


\section{Introduction}
The annihilation of an electron-positron pair into a pair of quarks
provides an ideal laboratory to test the theory of the strong interaction, Quantum
Chromodynamics (QCD)~\cite{Fritzsch:1973pi,Gross:1973ju,Gross:1973id}.
The free parameter of QCD, the strong coupling \as, can be determined
with events with more than two jets in the final state. To first order perturbation theory
the radiation of a gluon from a quark is proportional to \as. 
To determine this fundamental constant the observed three 
jet rate is compared to a perturbative expansion which predicts the
three jet rate as a function of a single parameter \as~\cite{Biebel:2001dm,Kluth:2006bw,Ellis:1991qj}. \par
Recently theoretical progress has been made leading to a significant
improvement in the prediction of the three jet
rate~\cite{GehrmannDeRidder:2007jk,GehrmannDeRidder:2007hr,GehrmannDeRidder:2008ug,Weinzierl:2008iv}
as a function of \as. Previously
\epem-event shape distributions and the three-jet rate were only known
to next-to-leading-order (NLO) accuracy, now QCD calculations to
next-to-next-to-leading-order (NNLO) are available. These
predictions were used to determine \as\ based on a single
value of the resolution parameter of the three-jet rate using the 
Durham algorithm~\cite{Catani:1991hj}, with data taken at
a centre-of-mass energy at 91 GeV with the ALEPH experiment~\cite{Dissertori:2009qa}.  \par
In this analysis we use data taken with the JADE experiment located at
the PETRA collider at DESY between the years 1979 and 1986. We
measure the strong coupling \as\ at six centre-of-mass-energies in the
range between 14 and 44~GeV.  Besides using a different energy range we
perform a fit  in a range  of the  three-jet resolution parameter.
We present a
matching scheme to combine  NNLO predictions together with 
next-to-leading logarithm approximations (NLLA). The matched
predictions are used to determine the strong coupling \as. 
The analysis follows closely the determination of \as\ measuring  the four-jet rate
using data collected with the JADE experiment~\cite{Schieck:2006tc}. 
\section{Observable}
\label{theory}
To determine a multijet-rate a jet finding algorithm has to be applied to 
particles observed in the final state. For this analysis we use 
the Durham
algorithm~\cite{Catani:1991hj}, which selects jets according to the
jet resolution parameter \ycut.
The Durham algorithm defines initially each particle
as a proto-jet and a resolution variable $y_{ij}$ is calculated for
each pair of proto-jets $i$ and $j$:
\be
 y_{ij}=\frac{2\mathrm{min}(E_i^2,E_j^2)}{E_{\mathrm{vis}}^2}(1-\cos\theta_{ij}),
\ee
where $E_{i}$ and  $E_{j}$ are the energies of jets $i$ and $j$,
$\cos\theta_{ij}$ is the cosine of the angle between them 
and $E_{\mathrm{vis}}$ is the sum of the  energies
of the detected particles in the event (or the
partons in a theoretical calculation). 
If the smallest
value of $y_{ij}$ is less than a predefined value $\ycut$, the pair
is replaced by a new proto-jet with four-momentum
$p_{k}^\mu =  p_i^\mu + p_j^\mu$, and the clustering starts again.
Clustering ends
when the smallest value of $y_{ij}$ is larger than $\ycut$, and the remaining
proto-jets are counted as final selected jets.\par
Perturbative QCD calculations predict the fraction of three jet events
$R_{3}(y_{\mathrm{cut}})$  as a function of \ycut\ and \as. 
The NNLO QCD prediction for the three-jet rate can be written as 
\be
R_{3}(y_{\mathrm{cut}})=\frac{\sigma_{\mathrm{\mbox{\scriptsize{3-jet}}}}(y_{\mathrm{cut}})}{\sigma_{\mbox{\scriptsize tot}}} \\
   = \ash \,  A_3(y_{\mathrm{cut}})+\ash^{2} \,
   B_3(y_{\mathrm{cut}})+\ash^{3} \, C_3(y_{\mathrm{cut}}),
 \label{NNLOcalc}        
\ee
with $\ash = \asmu/ (2\pi)$ the only free parameter. 
For this analysis the coefficients $A_3,\, B_3$ and $C_3$ are taken from~\cite{Weinzierl:2010cw}.
Equation~\ref{NNLOcalc} is shown for
renormalisation scale $\mu = Q$, where $Q$ is the physical scale
usually identified with the centre-of-mass energy $\rs$ for hadron
production in $\epem$-annihilation. The terms generated  by variation 
of the renormalisation scale parameter $\xmu=\mu/Q$ are implemented according
to~\cite{Weinzierl:2010cw}. \par
\begin{figure}[htb!]
\begin{tabular}[tbp]{cc}
\includegraphics[width=0.5\textwidth]{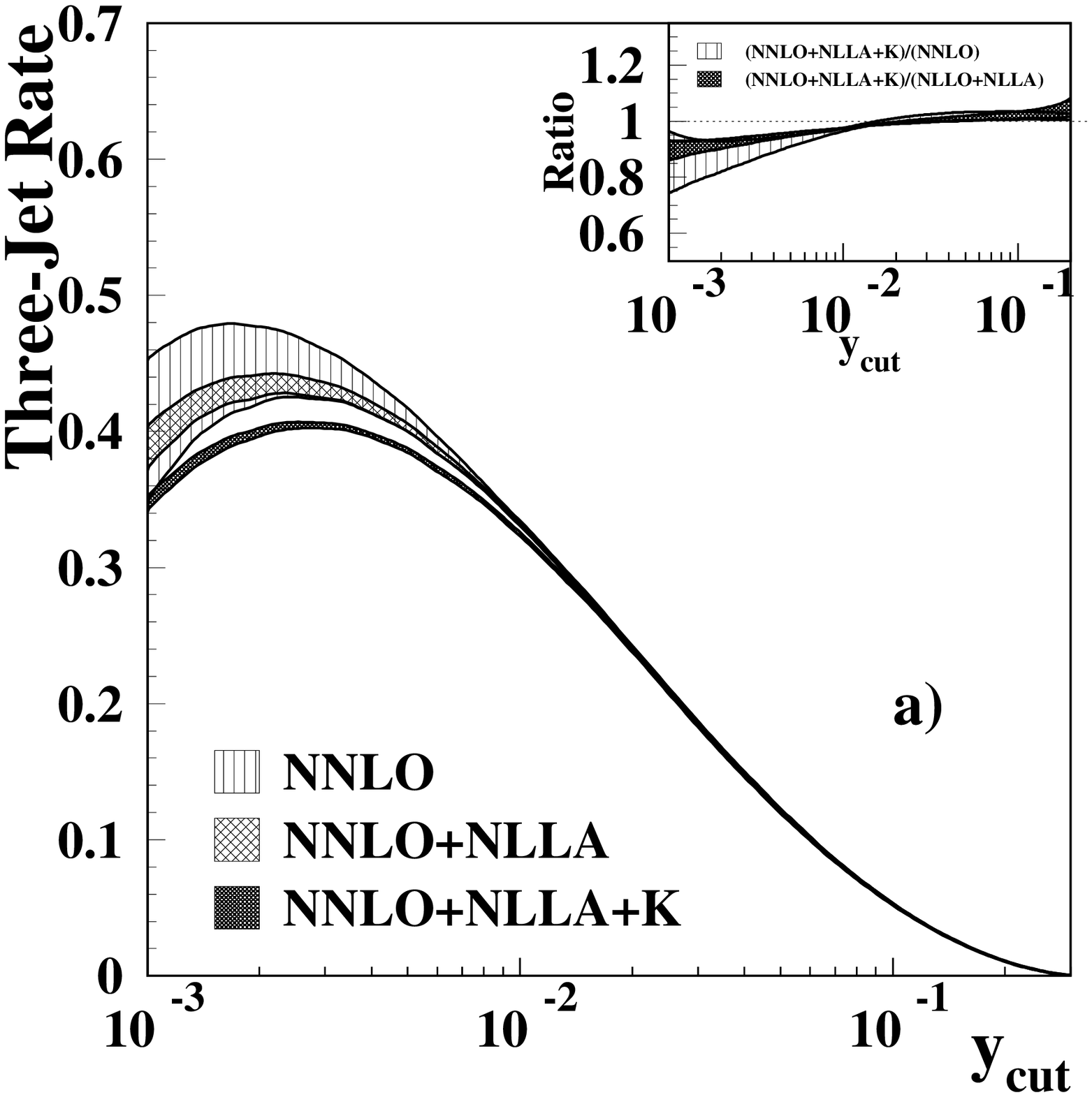} &
\includegraphics[width=0.5\textwidth]{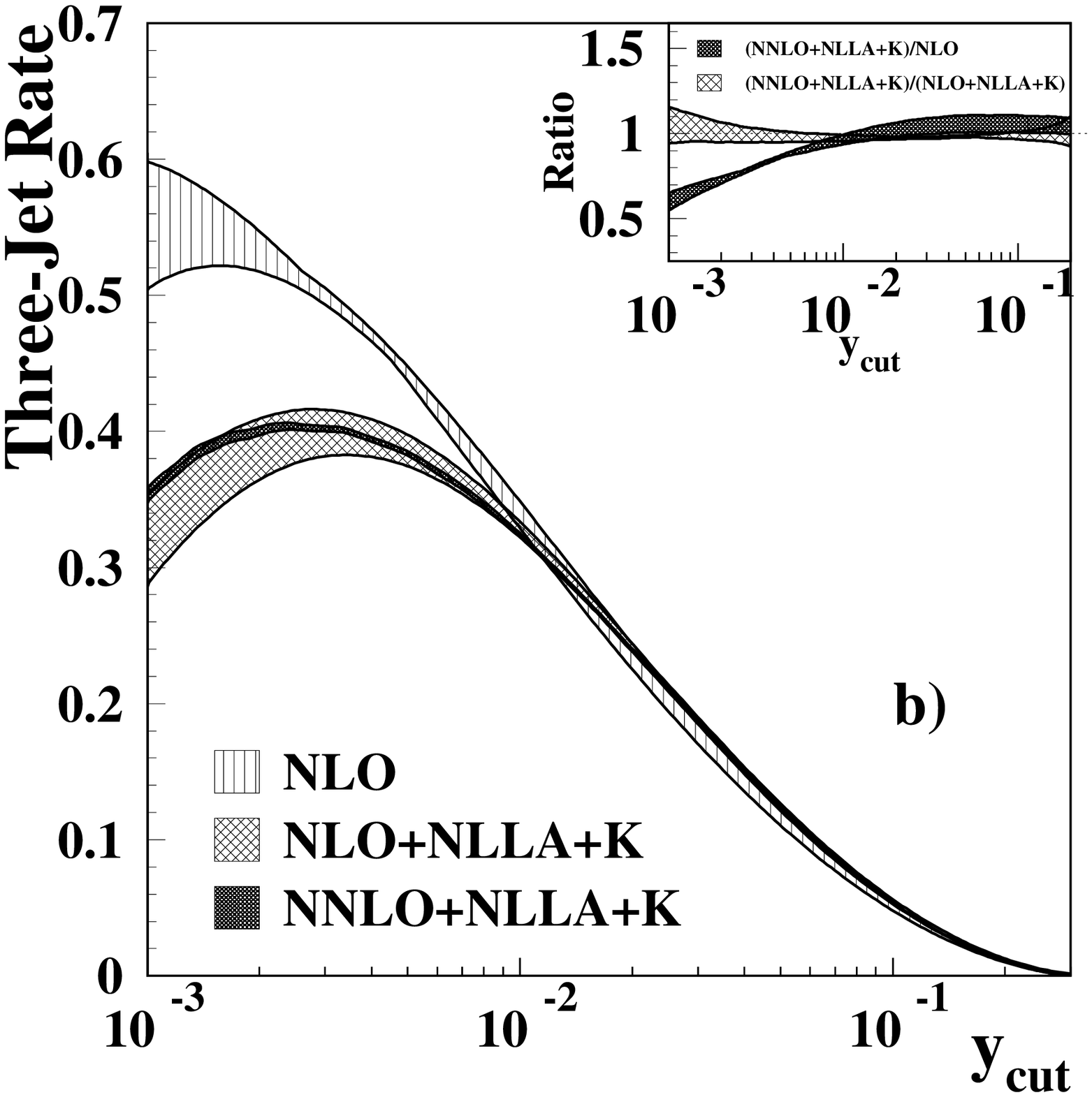}  
\end{tabular}
\caption{The QCD predictions are shown for the Durham three-jet rate calculated with a value 
of \asmz=0.1180 and at a centre-of-mass  energy of 35~GeV. In a) the NNLO prediction is compared to the matched 
NNLO+NLLA and NNLO+NLLA+K predictions. The insert shows the ratio between the different predictions. In b)  the matched 
and unmatched NLO predictions are compared to the matched NNLO prediction with the insert presenting the ratio to the fixed order
and the matched NLO predictions. For all calculations the uncertainty band reflects the uncertainty originating from setting 
the renormalisation scale factor $\xmu=\mu$/Q to 0.5 and 2.  }
\label{Fig:NNLOPrediction}
\end{figure}

It is well known that for small values of \yc\ the fixed order
perturbative prediction is not reliable, because the expansion
parameter $(\ash) L^2$, where $L = -\ln \yc$, logarithmically enhances the
higher-order corrections. For instance, $(\ash) \ln^2(0.01)\approx
\mathrm{O(1)}$. Thus, one has to perform the all-order resummation
of the leading and NLLA contributions.
This resummation is possible for the Durham algorithm using the coherent
branching formalism \cite{Catani:1991hj,Brown:1991hx}. Matched NLO+NLLA predictions
for jet rates have been compared to ALEPH data and good agreement was found, see Fig. 2 of ~\cite{Nagy:1998kw}.
There an improved resummation formula was used, where
part of the subleading logarithms, that can be controlled systematically
by including the `$K$-term'~\cite{Kodaira:1981nh,Davies:1984sp,Catani:1988vd}, are also taken into account (NNLO+NLLA+K). In this
paper we use the same resummation formula and extend the matching to the
NNLO accuracy, which requires the expansion of the improved resummation
formula up to O($\ash^3$). In the resummed prediction we use the one-loop
formula for the running coupling. One could also use higher-loop running,
but the difference would be in the coefficients of the subleading
logarithms (NNLL and higher), which are not controlled systematically
in the resummed prediction and therefore are neglected. Expanding the
resummation formula in \cite{Catani:1991hj}, we find
\begin{equation}
R_3 = \sum_{i=1}^3 \sum_{j=i}^{2i} (\CF \ash)^i R_3^{(i,j)} L^j
+ {\rm O}(\ash^4)\,,
\label{eq:R3NLLexpansion}
\end{equation}
with the $R_3^{(i,j)}$ coefficients are
given in Table~\ref{tab:R3NLLexpansion}. The dependence on the
renormalisation scale $\mu$ can be obtained by making the substitution
\begin{equation}
\ash \to \ash \sum_{n=0}^2 \left(\beta_{0} \ash \log\frac{\mu}{Q}\right)^n\,,
\label{eq:scaledep}
\end{equation}
and keeping the terms up to O($\ash^3$) in the expansion.
In Eq.~(\ref{eq:scaledep}),
\begin{equation}
\beta_{0} = \frac{11}{3}\CA - \frac{2}{3} \nf
\end{equation}
with \CA\ and \CF\ are the quadratic Casimir operators of the gauge
group in the adjoint and fundamental representation, while \Nf\ is the
number of light flavours (we use \Nf=5). The matched NNLO+NLLA+K
predictions compared to NNLO, NLO and matched NLO+NLLA+K, NNLO+NLLA 
predictions are shown in Fig.~\ref{Fig:NNLOPrediction}. \par
As opposed to event-shapes, such as the $y_{23}$-distribution~\cite{Akrawy:1989rg, Komamiya:1989hw}, jet
rates do not obey simple exponentiation (except for the two-jet rate).
For an observable that does not exponentiate, the viable matching
scheme is the so-called $R$-matching~\cite{Catani:1992ua}.  To obtain the $R$-matched predictions, we
subtract the expansion of $R_3$ from the resummation formula and add
the corresponding NNLO prediction given by Eq.2,
\begin{eqnarray}
R_3^{R-{\rm match}}(\ycut) & = &
R_3^{{\rm NLL}}(\ycut) +
\ash \Big(A_3(\ycut) - A^{{\rm NLL}}_3(\ycut)\Big)   \\ \nonumber 
 && + \ash^2 \Big(B_3(\ycut) - B^{{\rm NLL}}_3(\ycut)\Big) 
 + \ash^{3} \Big(C_3(\ycut) - C^{{\rm NLL}}_3(\ycut)\Big)
\,,
\end{eqnarray}
where
$A^{{\rm NLL}}_3 = \CF \sum_{j=1}^2 R_3^{(1,j)}L^j$,
$B^{{\rm NLL}}_3 = \CF^2 \sum_{j=2}^4 R_3^{(2,j)}L^j$,
$C^{{\rm NLL}}_3 = \CF^3 \sum_{j=3}^6 R_3^{(3,j)}L^j$ and
$A_3$, $B_3$, $C_3$ as in Eq.~\ref{NNLOcalc}.
Also in the case of jet rates the resummed logarithm is fixed to $L = -\ln  \ycut$
unambiguously, and the theoretical uncertainty of the
prediction is estimated by varying the renormalisation scale (see
Fig.~\ref{Fig:NNLOPrediction}) as in~\cite{Bethke:2008hf}. The theoretical 
uncertainty estimated by varying the renormalisation scale is found to 
be larger by using matched NNLO+NLLA+K predictions 
instead of NNLO+NLLA predictions.
\begin{table}[h]
\begin{center}
\begin{tabular}{|l||c|c|c|c|c|c|}
\hline
\hline
\lower.5ex \hbox{$i$}{\Large $\diagdown$}\raise.5ex \hbox{$j$}
\vlarge  & 1  & 2 & 3 & 4 & 5 & 6 \\ \hline \hline
\vlarge 1 & -3 & 1 &   &   &   &   \\ \hline
\vlarge 2 & &$-9-\frac32 b_0 + K$&$6+\frac12(b_0 + x)$&$-1-\frac{x}{12}$& &
\\ \hline
\vlarge  & &  & $-\frac{27}2$
& $+\frac{27}2 + \left(\frac32 +\frac38 b_0\right) x$
& $-\frac92$
& $\frac12$ \\
\vlarge 3 & &  & $-\left(\frac{27}4+\frac58 b_0\right) b_0$
& $+\left(\frac{19}{4} + \frac5{24} b_0\right) b_0$
& $-\left(\frac{5}6+\frac7{120} x\right) b_0$
& $+\frac{x}{12}$ \\
\vlarge  & &  & $+\left(6 + \frac56 b_0 + \frac12 x\right) K$
& $- \left(2 + \frac{x}6\right) K$
& $-  \left(\frac34 + \frac{3}{40}x\right)x$ & $+\frac{x^2}{120}$ \\
\hline
\hline
\end{tabular}
\end{center}
\caption{\label{tab:R3NLLexpansion}
Coefficients $R_3^{(i,j)}$ of the expansion in Eq.~(\ref{eq:R3NLLexpansion}) with 
$K=\left(\frac{67}{18}-\frac{\pi^2}{6}\right)x-\frac{10}{9}\yf$, $x = \frac{\CA}{\CF}$, $\yf = \frac{\Nf}{2C_F}$
and $b_0 = \beta_{0}/\CF$.}
\end{table}
\section{Analysis Procedure}
\subsection{The JADE Detector}
\label{TheJadeDetector}
A brief description of the JADE detector focusing on the major
detector parts used in this analysis is given here. Primarily the
momenta of charged and neutral particles are used in this analysis. The trajectories
of charged particles are mainly reconstructed with the central tracking detector,
consisting mainly of a large scale jet chamber.
The tracking detector is located within a 0.48 T
solenoidal magnetic field. 
Charged and neutral particles except muons and neutrinos are
reconstructed with an electromagnetic calorimeter which surrounds the
magnetic coil. The calorimeter consists of lead glass Cherenkov counters and is
separated in a barrel and two endcap regions. A more complete summary
of the JADE detector can be found in~\cite{Naroska:1986si}.
\subsection{Data and Monte Carlo Samples}
\label{MC}
For this analysis we use the identical data sample as 
used for previous JADE analyses~\cite{Schieck:2006tc,Bethke:2008hf,Pahl:2008uc}. 
The data were taken between 1979
and 1986, adding up to an integrated luminosity of about 195~\invpb. 
It is subdivided into six data taking periods with different average
centre-of-mass energies of 14~GeV, 22~GeV, 34.6~GeV, 35~GeV, 38.3~GeV
and 43.8~GeV. The number of selected events ranges between 1403
events  at 22~GeV and 20876 events at 35~GeV. \par
To correct for acceptance and resolution effects as well as for
hadronisation effects a large sample of Monte Carlo events is
generated.  The Monte Carlo generators are tuned to match 
events taken with the OPAL experiment at a centre-of-mass 
energy of 91~GeV~\cite{Alexander:1995bk,Abbiendi:2003ri}.
Using these same parameters, except of setting the appropriate 
centre-of-mass energy, a good description of the JADE data, down
to smallest energies, is achieved~\cite{pedrophd}. 
For the correction of acceptance and 
resolution effects the events are passed through a simulation of the
JADE experiment.  Events generated with \py~5.7~\cite{Sjostrand:1993yb}
are used as default for the correction of acceptance and
resolution effects. As a cross check events 
simulated with the \hw~5.9~\cite{Marchesini:1991ch} event
generator are utilised.
To assess the changes in the three-jet rate distribution originating from the 
transition from partons to hadrons three different Monte
Carlo generators are applied. As default \py~6.158 is used and events produced
with the \hw~6.2 and \ar~4.11~\cite{Lonnblad:1992tz} generators are used
as a consistency check. \par
Due to the good agreement between data and the predictions from
the \py, \hw\ and \ar\ Monte Carlo event generators (see section~\ref{Text:JetRates}) we do not
utilise event generators which incorporate high-multiplicity matrix
elements which were not yet tuned to our purpose. 
This choice is justified by the fact that 
generators used in our study use leading-order matrix elements combined with a 
leading-logarithm parton shower which do provide a satisfactory
description of the three parton final states studied in this analysis
as shown in~\cite{Bethke:2008hf,Abbiendi:2011nnlo}. 
\subsection{Event Selection}
The measurement of the strong coupling \as\ is based on the
analysis of well measured hadronic events. A detailed
description of the hadronic event selection can be found in~\cite{Schieck:2006tc}.
Cuts applied to the events are based on the number of charged tracks, total visible energy and
momentum imbalance.  The cut on the momentum imbalance reduces events  
emitting a high energetic photon in the initial state, leading to a
reduced hadronic centre-of-mass energy.
The cuts on the number of charged tracks and total visible energy minimise to an insignificant
level the number of events from hadronic tau decays and from hadronic final states 
originating from two-photon scattering.
\subsection{Corrections to the Data}
\label{DataCorrection}
The corrections applied to the data follow exactly the same procedure
as summarised in~\cite{Schieck:2006tc}. For the calculation of the
three-jet rate all charged tracks and electromagnetic clusters are
considered. 
The estimated minimum ionizing energy from tracks associated with
electromagnetic calorimeter clusters is subtracted from the cluster
energies.
For the correction procedure two different
categories of  jet rate distributions are defined for simulated
events, the so-called {\em detector level} and 
the {\em hadron level} distribution. 
The {\em detector level} distribution of simulated events is obtained by using all selected
tracks of charged particles and electromagnetic clusters.
The {\em hadron level} distribution is obtained by using the true four-momenta of stable particles, where particles with a
lifetime of $\tau > 300$~ps are declared as stable. 
Simulated events with photon initial state radiation (ISR) leading to the
centre-of-mass energy reduced by more than 0.15 GeV are rejected from the {\em hadron level}\,\footnote{The cut value 
of 0.15 GeV is purely technical and corresponds to a clean separation of the soft and hard ISR events in the simulation.}.
Thus the correction for experimental effects explained below also takes care of residual effects due to initial state radiation.
Only hadronic events originating from the primary production of  u,d,s or c quarks are
considered. \par 
Before correcting the three-jet rate data distribution the expected
contribution from $\epem \to \bbbar$-events as expected from simulated events
 at the
{\em detector level} is subtracted from the three jet-rate. 
About 1/11 of all \qqbar-events are \bbbar-events and the expected number of \bbbar-events is 
subtracted from the observed number of data events at each \ycut-bin.
The hadronic events used in this analysis correspond to events with \epem\ annihilating to a pair of
u,d,s or c-quarks.
The distribution corrected for $\epem \to \bbbar$-events is then multiplied bin-by-bin with the ratio
of the {\em hadron level} distribution divided by {\em detector level}
distribution. A correction method based on a matrix unfolding method returns compatible results
within statistical uncertainties~\cite{pedrophd}.
The impact on the measurement due to changes of the $b$-quark fragmentation
in the simulation are covered by the systematic uncertainty (see section \ref{systematic}) assigned
to the correction for \bbbar\-events ~\cite{pedrophd} .
The numerical results of the corrected distributions are
summarised in Tables~\ref{hadron_tab_1} and~\ref{hadron_tab_2}. 
\section{Systematic Uncertainties}
\label{systematic}
In order to assess the systematic uncertainty of the measurement of
the strong coupling \as\ we evaluate several possible sources. For
each variation the difference to the result with respect to the
default analysis procedure is taken as a contribution
to the systematic uncertainty. The systematic uncertainty is assumed
to be symmetric around the default value. No systematic uncertainty 
is evaluated related to the fact that massless theoretical predictions
are used since the contribution from $\epem \to \bbbar$-events  
is subtracted from the data distribution (see section~\ref{DataCorrection}).
The uncertainty originating from correcting for ISR effects is 
small and no systematic uncertainty is assigned.
\subsection{Experimental Uncertainties}
The assessment of the experimental uncertainty follows exactly the
procedure described in detail in~\cite{Schieck:2006tc}.  The analysis is repeated
with a slightly modified event and track selection, using a different reconstruction
software, using different MC models for the correction of detector effects 
and modified fit ranges.
In addition the amount of subtracted $\epem \to \bbbar$-events is modified
by $\pm5\%$.
The differences between the results obtained from the modified fits and the default fits
are added in quadrature and taken
as the combined experimental systematic uncertainty. The main
contribution originates from using a different version of the 
reconstruction software and using \hw\ instead of \py\ for the correction of detector effects.
\subsection{Hadronisation}
The standard analysis uses \py\ to evaluate the change of the
three-jet rate distribution originating from the transition from
partons to stable hadrons (see section~\ref{MC}). Only 
the process $\epem$ to a pair of u,d,s or c-quarks is simulated.
To assess the
uncertainty the fit is repeated with alternative Monte Carlo
models. 
For this we use \hw\ and \ar\ instead of \py\ and the
difference to the standard \py\ correction is taken as systematic
uncertainty. In all cases the larger difference is seen by using the
\hw\ Monte Carlo simulation instead of \py. \par
A variation of the parameters describing the hadronization model
leads to significant smaller changes on the measurement of the strong
coupling \as\ than applying a different hadronization scheme, like
\hw~\cite{pedrophd}. For this reason we quote only the largest uncertainty coming
from using a different hadronization model.
As described in ~\ref{MC} we use different versions of the Monte Carlo generator for
detector and hadronisation corrections. However, the
hadronisation corrections for each generator are consistent 
within their statistical uncertainty.
\subsection{Theoretical Uncertainties}
The theoretical prediction of the three-jet rate distribution 
is  a truncated 
asymptotic power series. The uncertainties originating from 
missing higher order terms are assessed by varying 
the renormalisation scale parameter $\xmu = \mu / \rs$. For this \xmu\ is set to
two and to $0.5$. The larger deviation from the fit using the default
setting $\xmu = 1$ is taken as systematic uncertainty. 
We assess the theoretical uncertainties by applying
NNLO+NLLA+K QCD predictions in the fit, since the 
uncertainties using NNLO+NLLA predictions 
without $K$-term are found to be smaller.
Effects from electroweak corrections are neglected. 
\section{Results}
\subsection{Three-Jet Rate Distributions}
\label{Text:JetRates}
\begin{figure}[!htb]
\begin{center}
\begin{tabular}[tbp]{cc}
\includegraphics[width=0.38\textwidth]{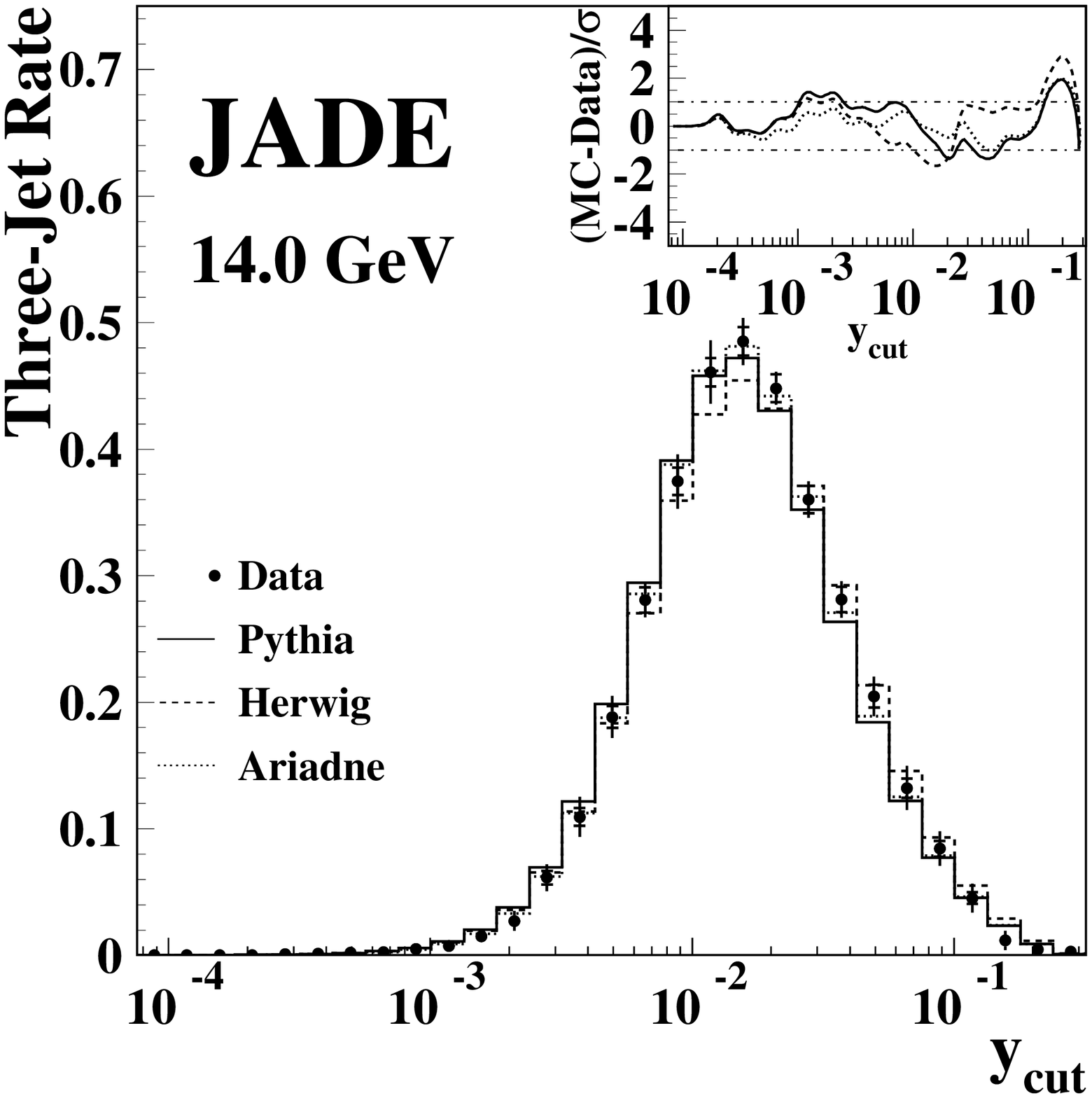} &
\includegraphics[width=0.38\textwidth]{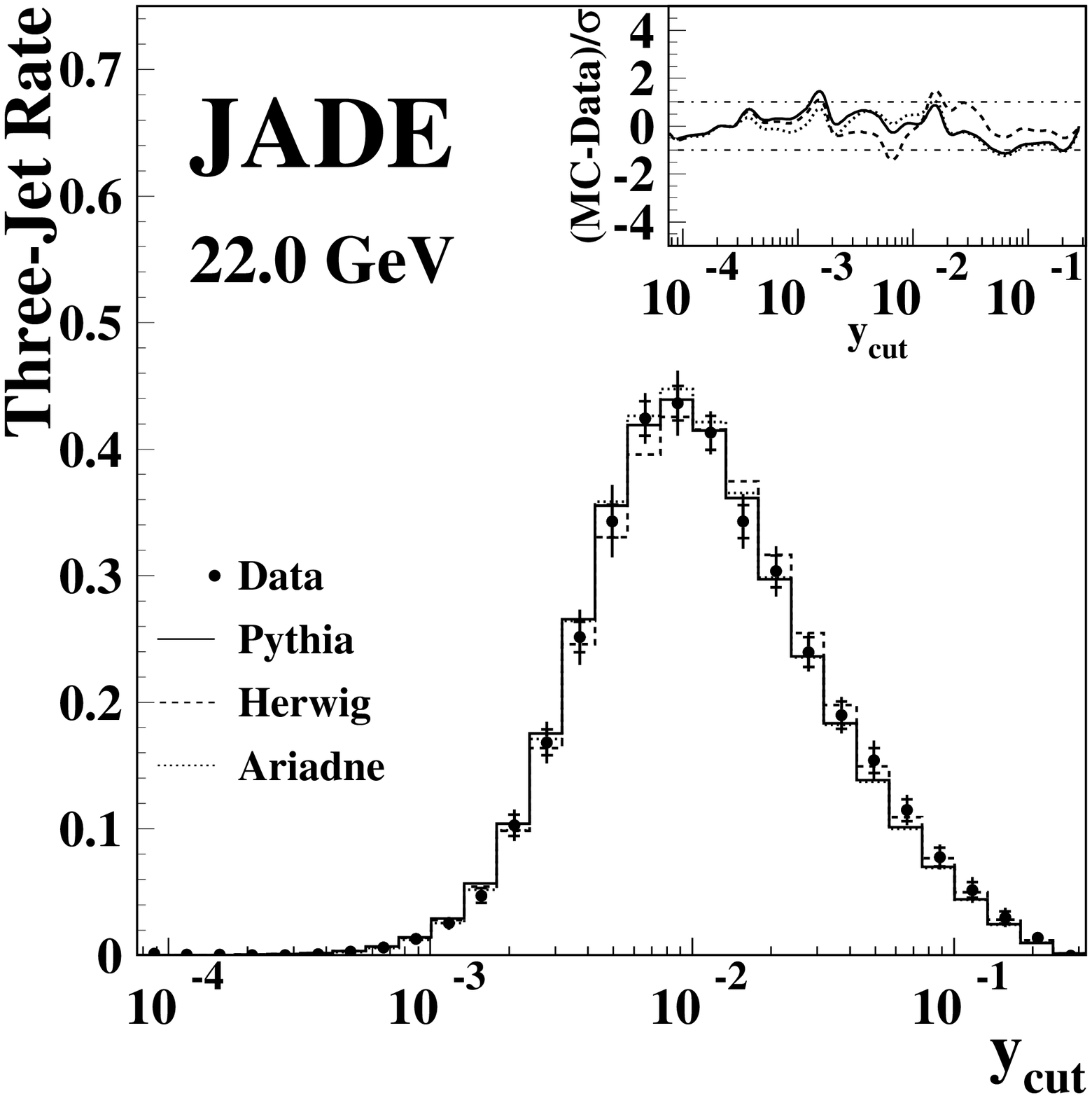}  \\
\includegraphics[width=0.38\textwidth]{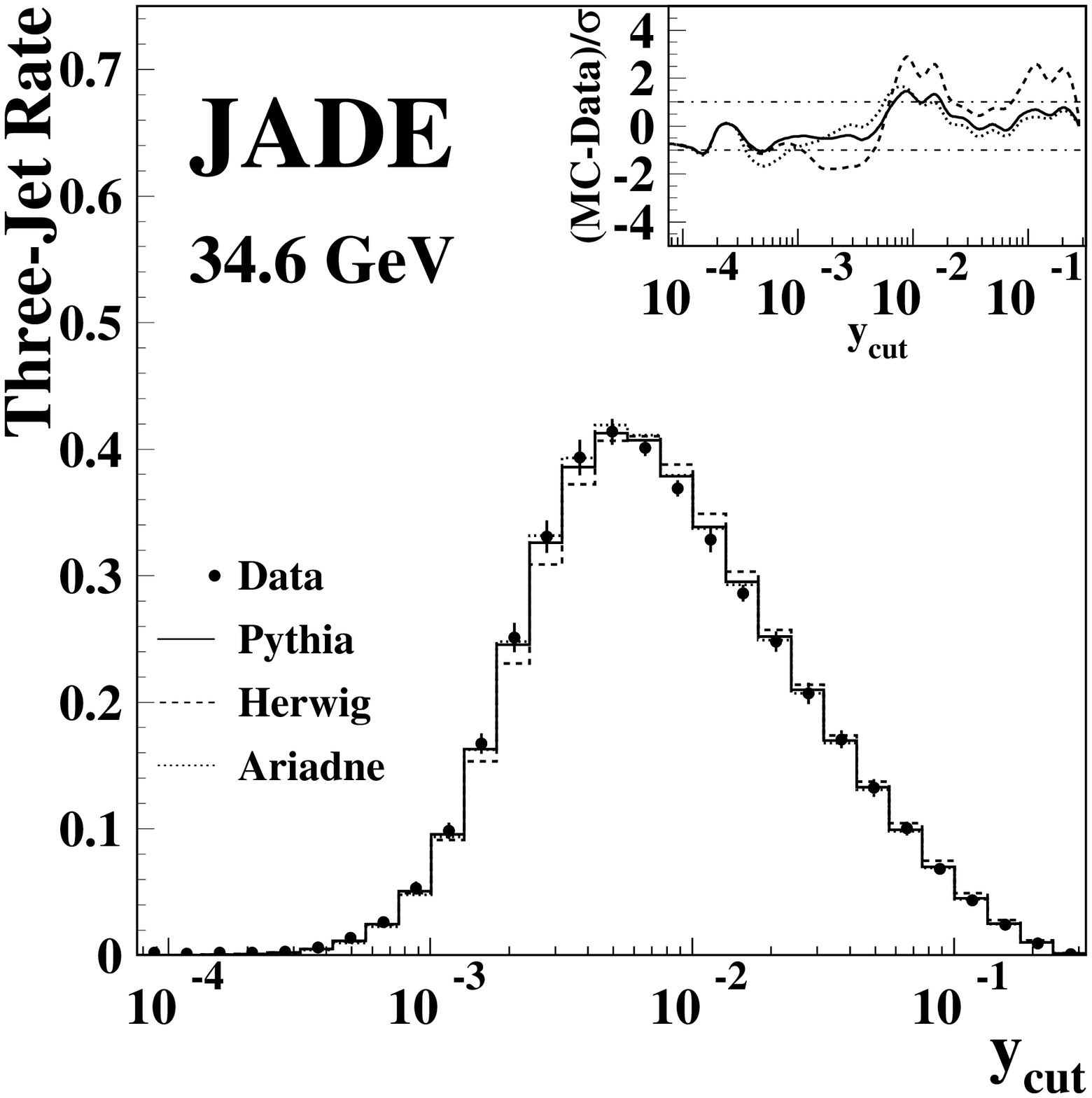} &
\includegraphics[width=0.38\textwidth]{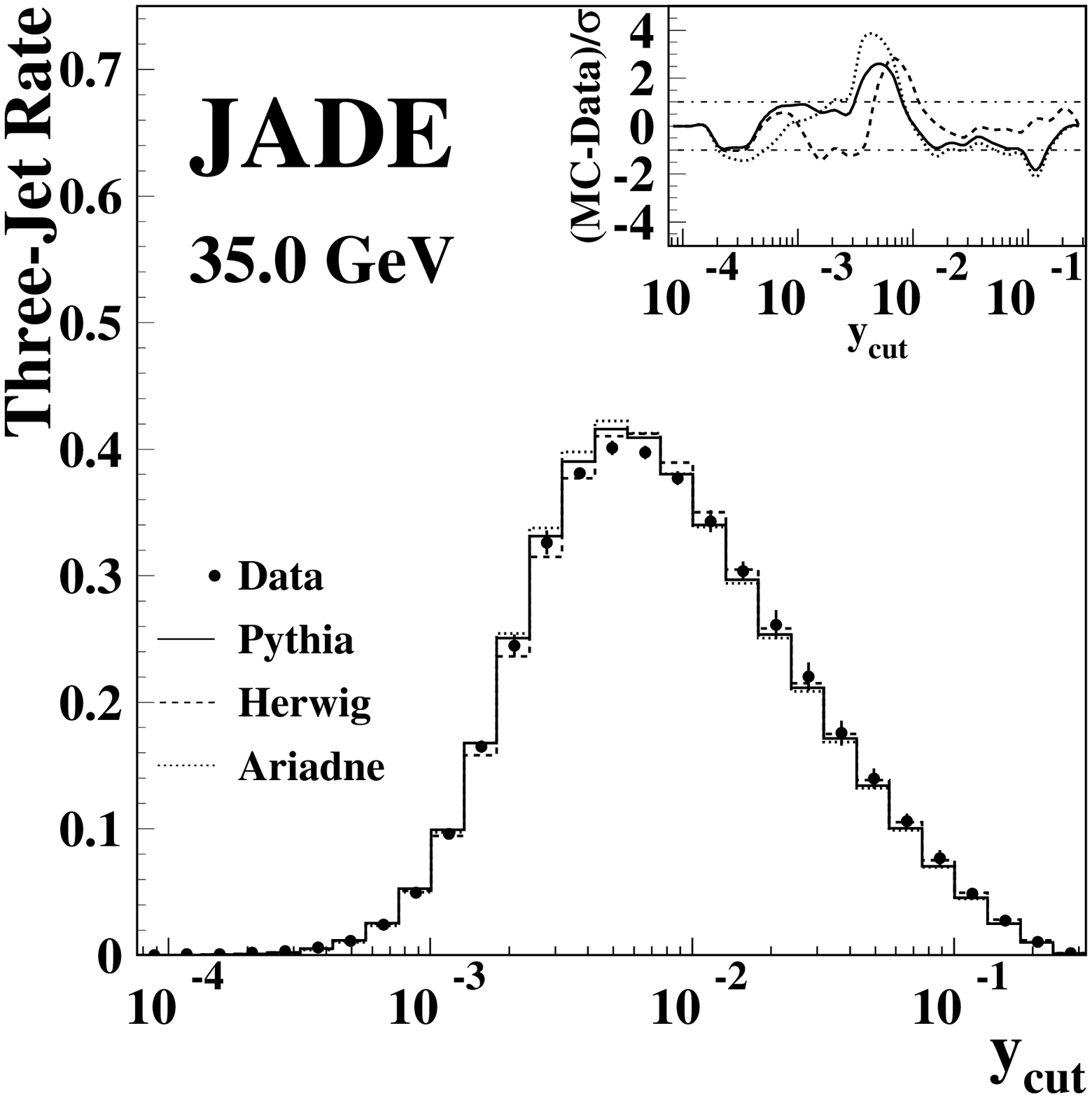}  \\
\includegraphics[width=0.38\textwidth]{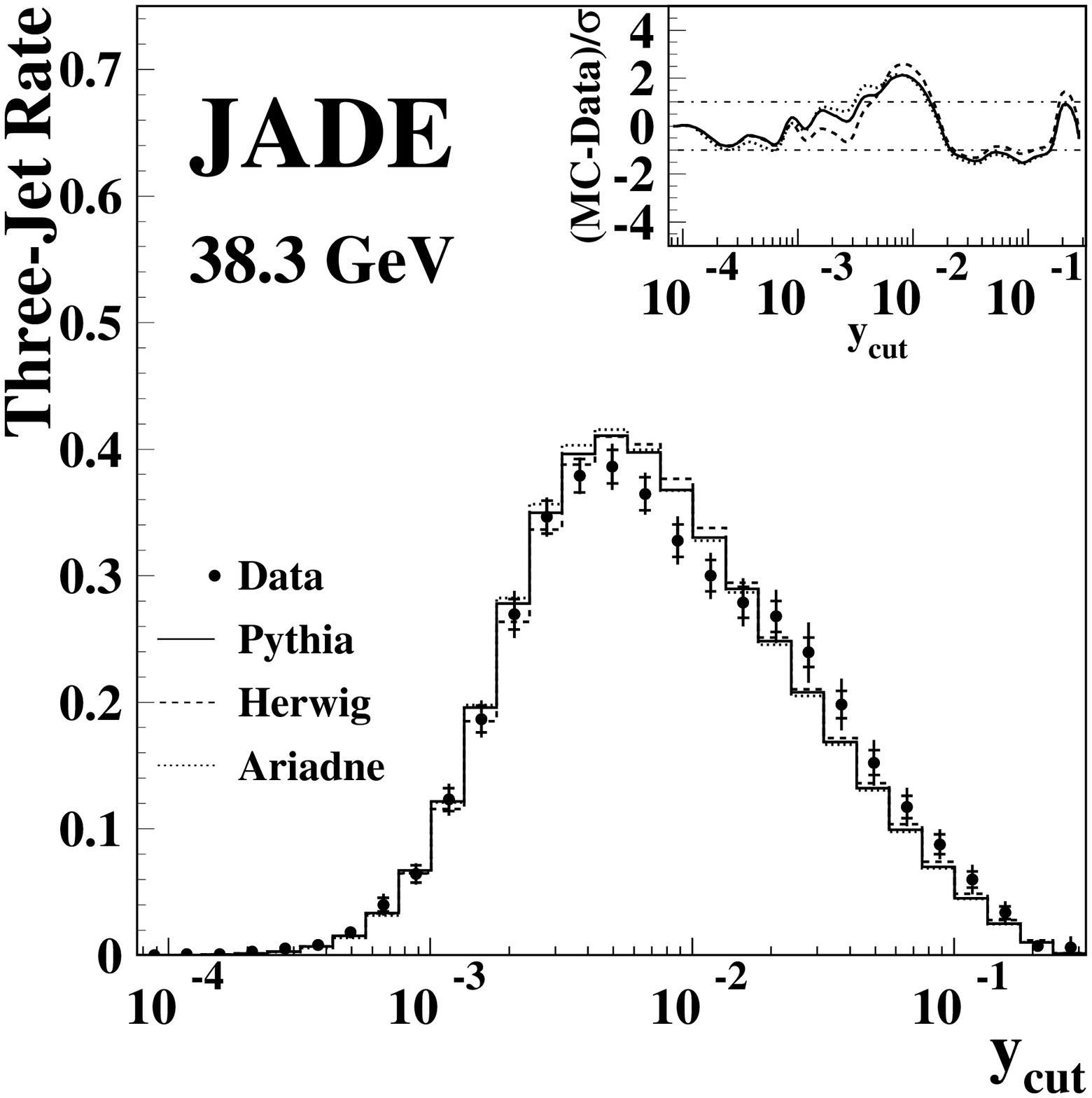} &
\includegraphics[width=0.38\textwidth]{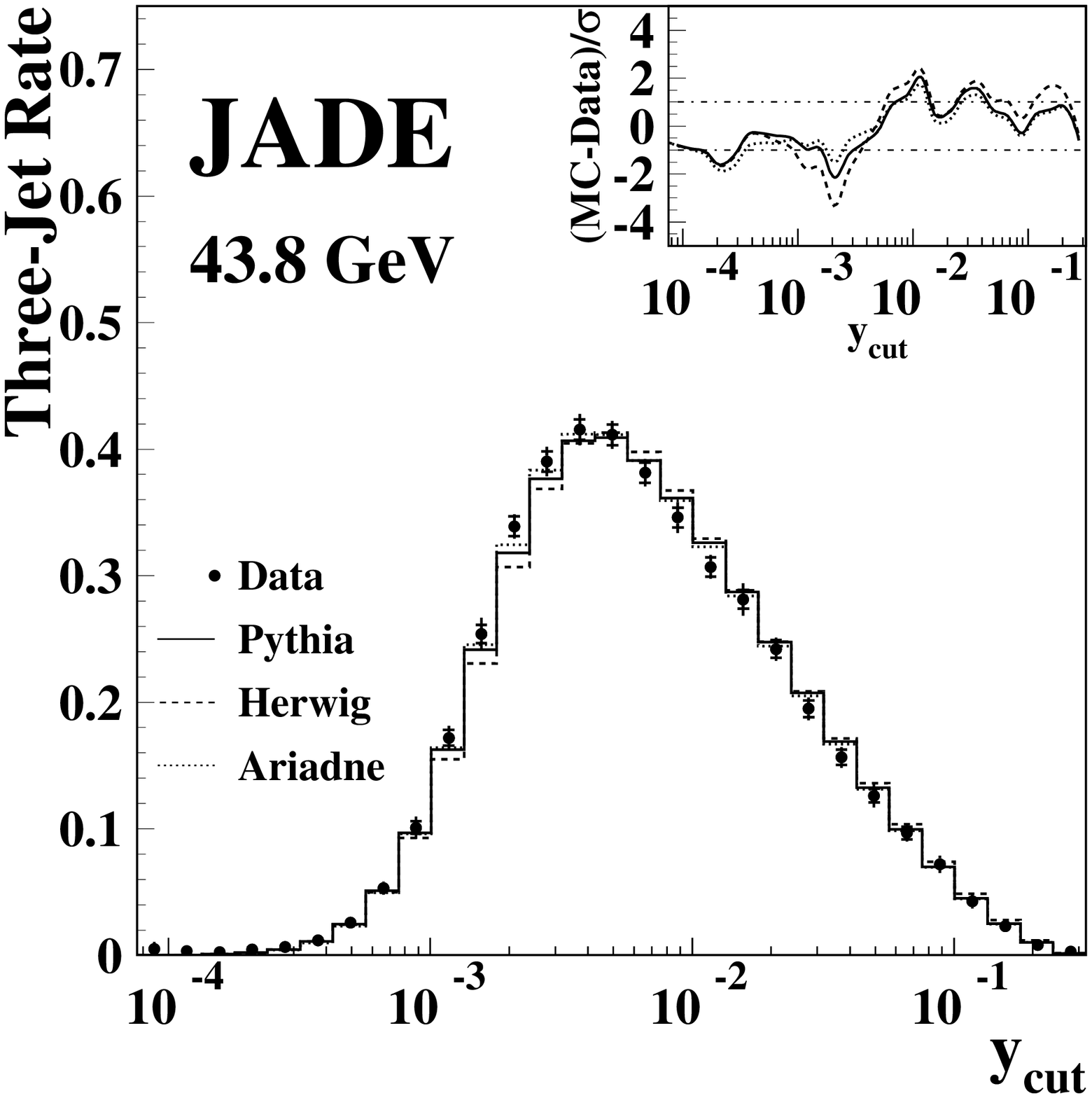} \\
\end{tabular}
\caption{The three-jet rate as a function of the resolution
  parameter \ycut\ is shown  for 
  centre-of-mass energies \rs\ of 14.0 to 43.8~GeV. The
  data distributions are corrected for resolution and acceptance effects from the
  detector and the contributions from  $\epem \to \bbbar$-events are subtracted. Data
  points are shown with statistical uncertainty (inner part) and combined
  statistical and experimental uncertainty. The histograms show the
  comparison to the predictions obtained with Monte Carlos simulation
  using \py, \hw\ and \ar. The inserts show the deviation from the
  simulated distribution normalised to the combined statistical and
  experimental uncertainty.}
\label{hadron}
\end{center}
\end{figure}

The three-jet rates as a function of \ycut\  at
\rs = 14, 22, 34.6, 35, 38.3 and 43.8 GeV are shown
in Fig.~\ref{hadron}. The rates are corrected for resolution and acceptance effects.
The estimated
contributions from $\epem \to \bbbar$-events are subtracted.  The
distributions are compared with the predictions obtained from Monte
Carlo models used in this analysis. All models reproduce the
data distribution well. 
Most of the data points are within the one sigma
uncertainty band and a couple of points show a deviation of 
up to two sigma uncertainty.
There is an
apparent deviation of the simulation from the data in Fig.~\ref{hadron}
(bottom left) for \rs=38.3 GeV.  This is due to the positive correlations between
the data points which implies that a fluctuation in one data point is
also visible in the neighboring data points.
For additional information inserts 
show the deviation of the three-jet rate obtained from Monte Carlo with respect to the   
data, normalised to the statistical and experimental uncertainty
added in quadrature\,\footnote{Please note that correlations between
  the points are present and not taken into account for the insert plot.}.
\clearpage
\subsection{Determination of \boldmath{\as}}
\label{fitprocedure}
The value of the strong coupling is determined by a minimum-$\chi^{2}$
fit of the matched NNLO+NLLA+K predictions to the corrected data distributions,
separately for each centre-of-mass energy. The reduced renormalisation scale parameter
$\xmu=\mu  / \sqrt{s}$ in the theoretical predictions is set to the natural choice
\xmu=1. The QCD predictions describe the three-jet rate at parton
level only.
To correct for hadronisation  effects the matched QCD predictions
are multiplied  at each \ycut\ point by the ratio of the {\em hadron level}
distribution divided by the {\em parton level} distribution obtained
from simulated events. 
The {\em parton level} distribution in
simulation is obtained
from the final state partons after the parton shower has terminated, i.e. just before the 
hadronisation step. The {\em
hadron level} is defined in an identical way as described in section~\ref{DataCorrection},
containing only hadronic events with $\epem$ annihilating to a pair of u,d,s or c-quarks.
The ratio between the {\em hadron level} divided by the {\em parton level} estimated with simulated 
events is shown in the appendix  in Fig.~\ref{Hadron_Correction}.
Using the QCD-prediction corrected for hadronisation effects and the
corrected data distribution a $\chi^{2}$-value is calculated 
\be
 \chi^{2} = \sum_{i,j}^{n}(R_{3,i}-R(\as)_{3,i}^{\mathrm{theo}})
            (V(\mathrm{R_{3}})^{-1})_{ij}(R_{3,j}-R(\as)_{3,j}^{\mathrm{theo}})
\ee
with $i$ and $j$ being the \ycut\ points in the chosen
fit range and the $R(\as)_{3,i}^{\mathrm{theo}}$ values are the predicted
values of the three-jet rate.  
Each event can contribute to several points in the three-jet rate
distribution leading to correlation between different \ycut\ points. For this reason the
covariance matrix $V_{ij}$ is not diagonal and the off-diagonal
elements have to be computed.
We follow the approach described in~\cite{Schieck:2006tc} using 1000
subsamples with 1000 simulated events each. \par
When choosing the fit range, we took the following considerations into account.
The corrections
applied to the data (see section~\ref{DataCorrection}) reverting the 
imperfectness of detector and the correction of the QCD-predictions
due to  hadronisation effects  are required to be small.
In addition we require the leading log contribution in the
low \ycut\ region of the three-jet rate distribution to be well below
unity to ensure that the NLLA is valid. The leading log term is proportional to
$\ash\cdot \ln^{2}(1 / \ycut)$ and requiring  this term to
be well below unity leads us to a lower
limit of \ycut= 0.01 assuming a value of the strong coupling \asmz=0.118. 
The upper limit is determined by requiring  the leading order
contribution $A(\ycut)$ to be larger than zero.
The corrections are small in this range with the detector corrections being less than 
 $30\%$ and the hadronisation corrections being less than $30\%$, apart from the 
 corrections at the centre-of-mass energy of 14~GeV, which are up to $70\%$.
 The considerations described above lead to a fit range from 0.01 to
0.2, which is identical to the fit range
used for the determination of \as\ using the differential \yy\ event shape distribution~\cite{Bethke:2008hf}. \par
\begin{figure}[h!]
\begin{center}
\begin{tabular}[tbp]{cc}
\includegraphics[width=0.38\textwidth]{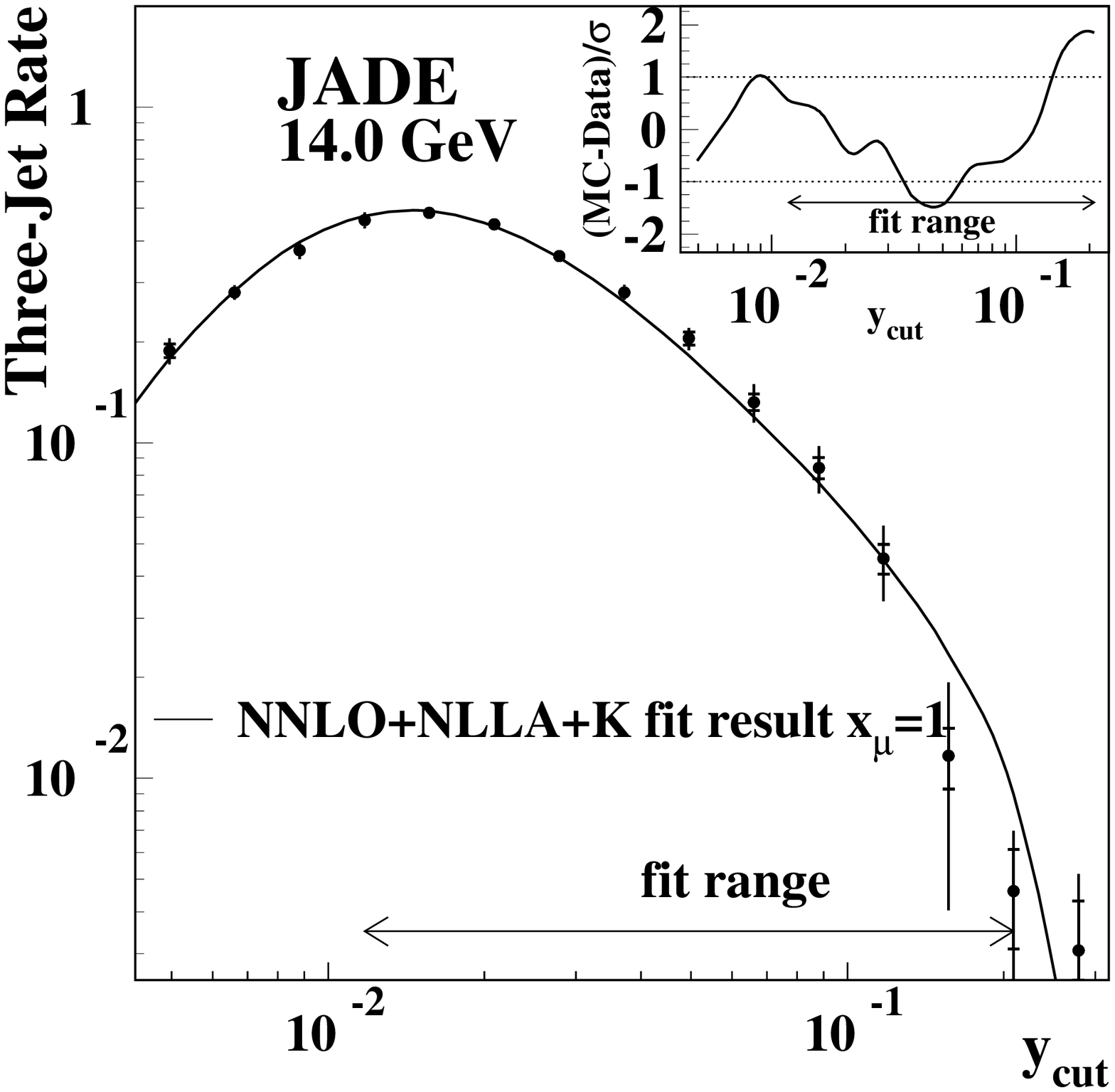} &
\includegraphics[width=0.384\textwidth]{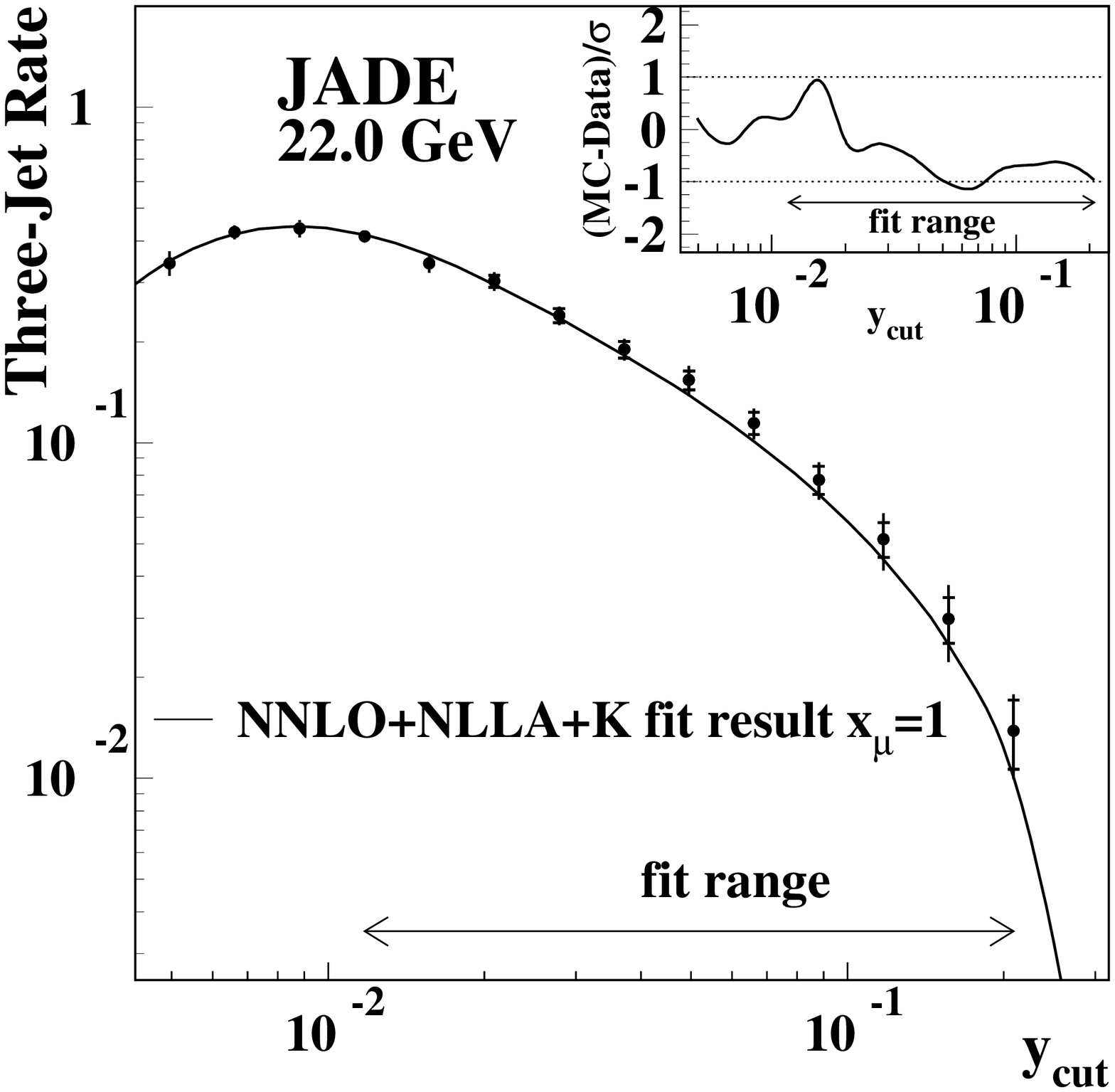}  \\
\includegraphics[width=0.38\textwidth]{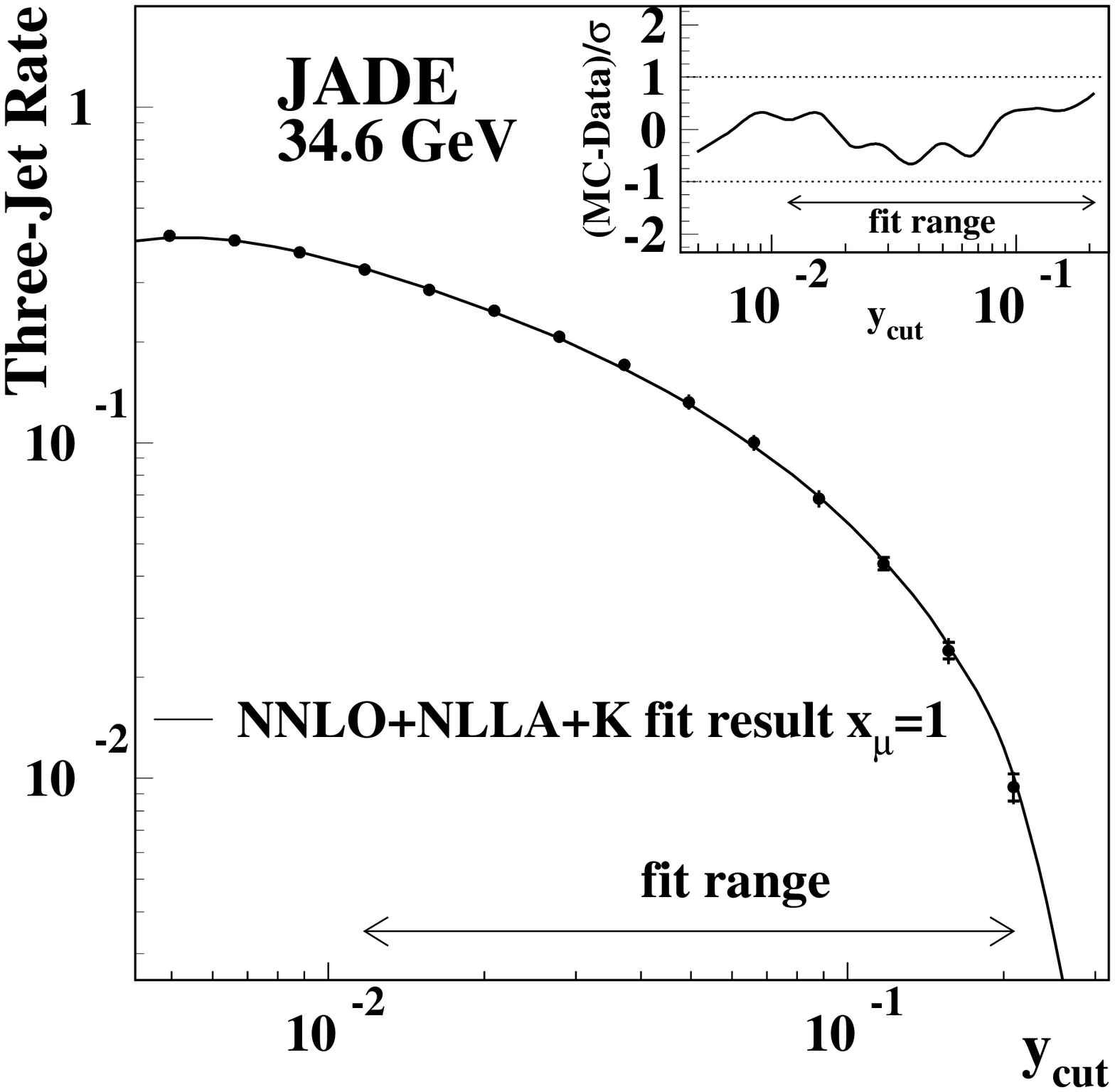} &
\includegraphics[width=0.38\textwidth]{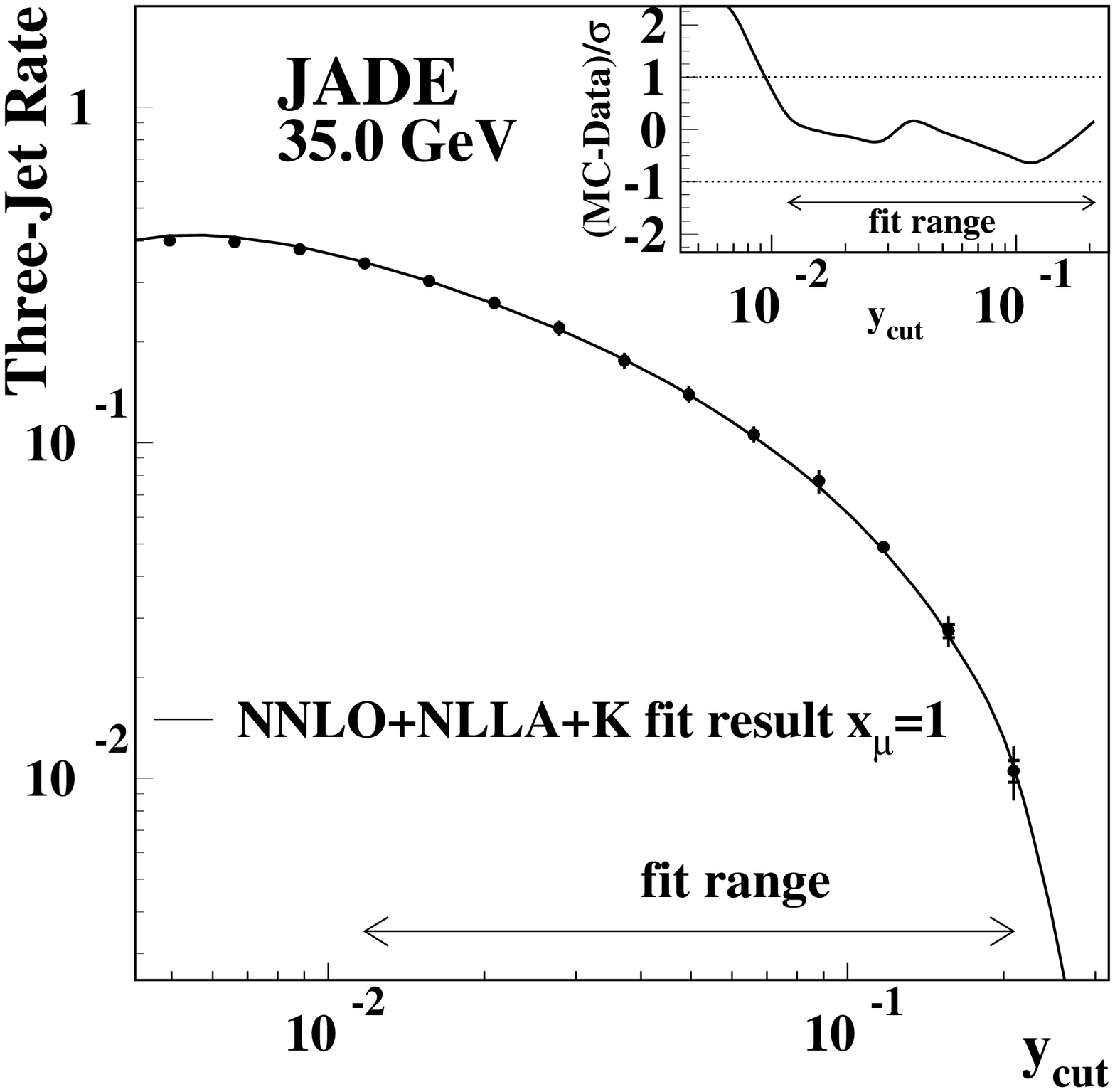}  \\
\includegraphics[width=0.38\textwidth]{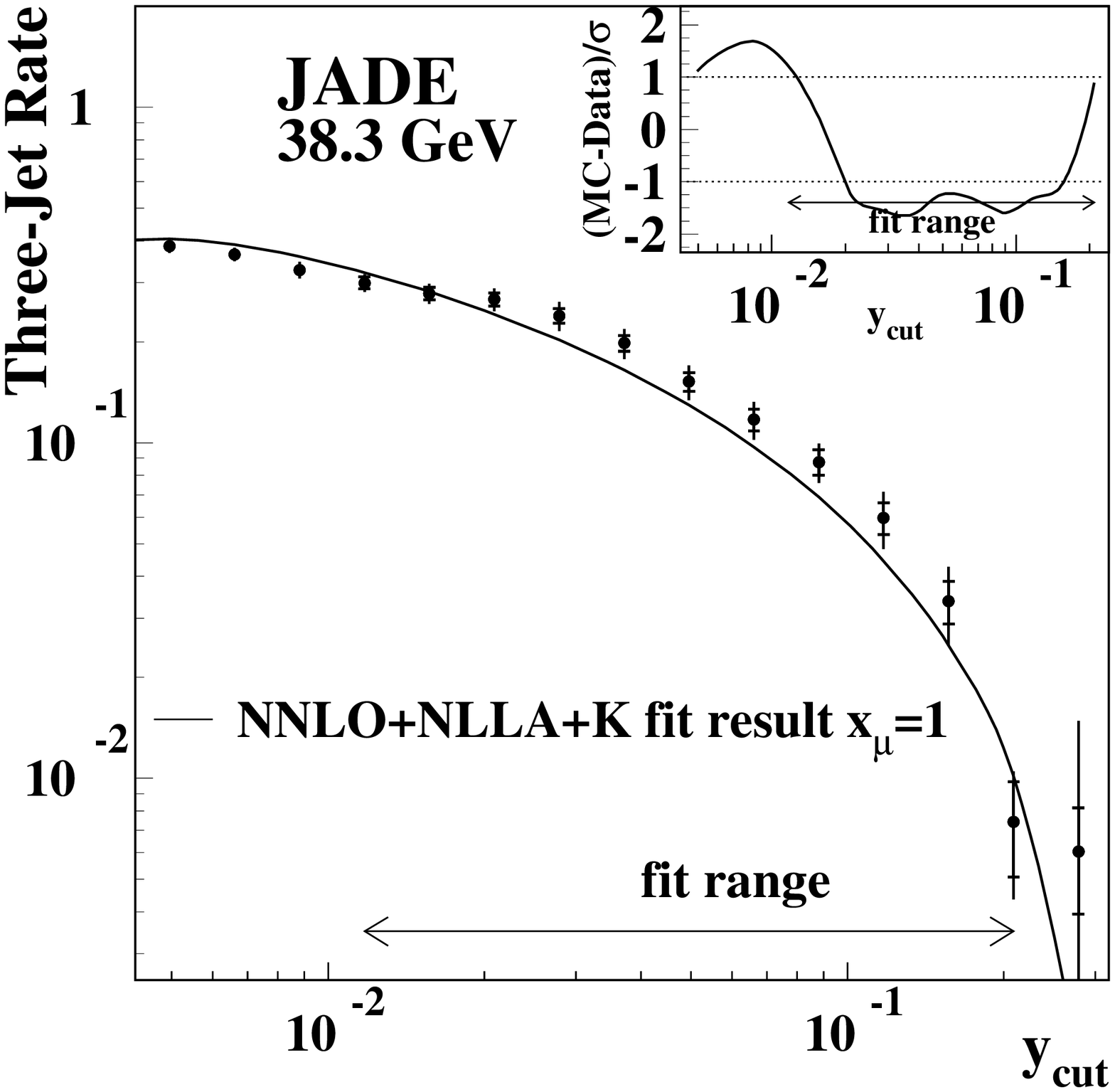} &
\includegraphics[width=0.38\textwidth]{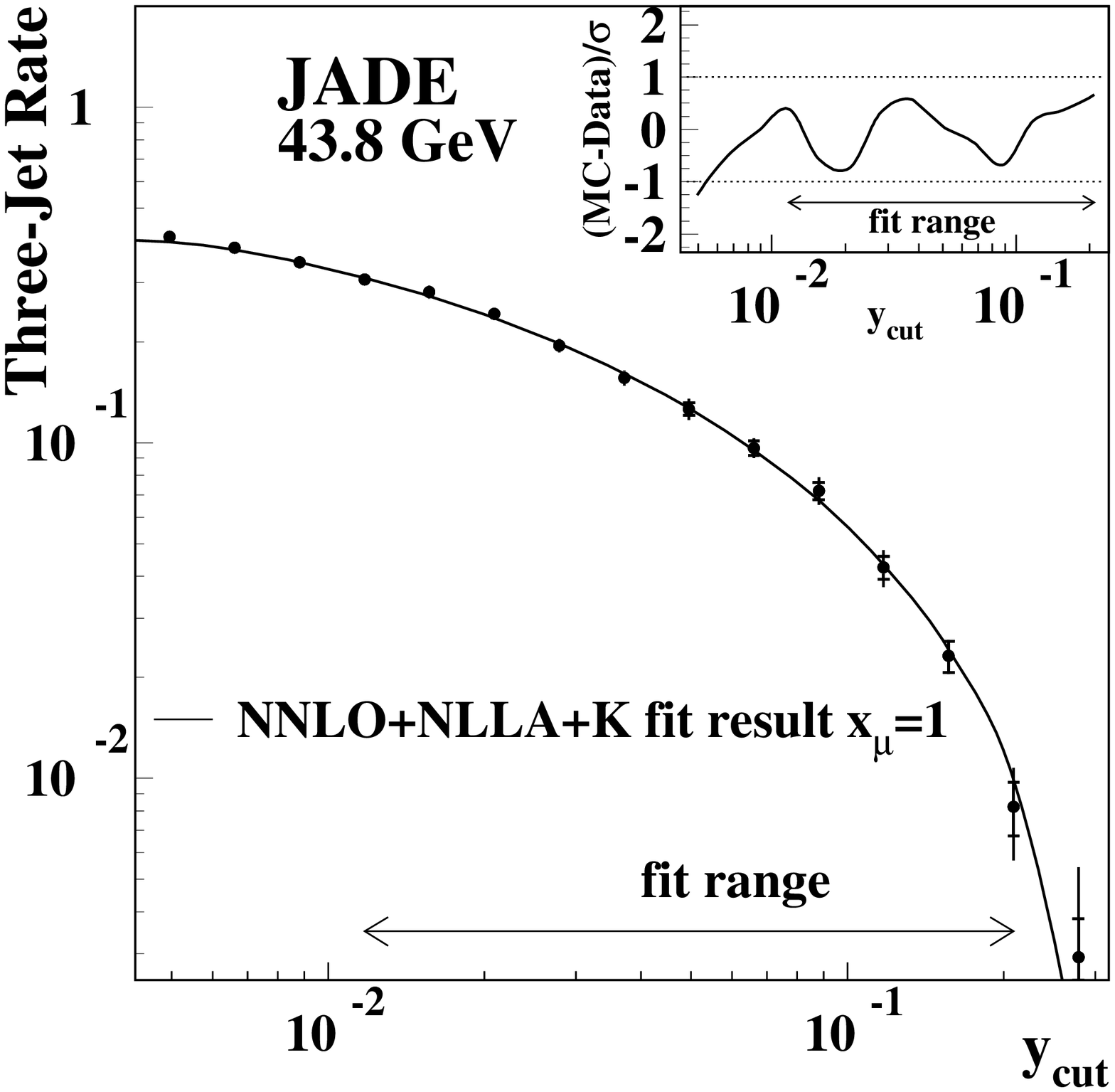}  \\
\end{tabular}
\caption{
The fit results at  centre-of-mass energies from 14.0 to 43.8~GeV are shown.
The inserts show the deviation of the data points from the
QCD-prediction with the \as-value obtained from the fit, normalised
to the combined statistical and experimental error.
}
\label{fit}
\end{center}
\end{figure}
\clearpage
The results of the fits to the three-jet rate distribution at
the various centre-of-mass energies are shown in Fig.~\ref{fit}.
The numerical results of the fits are summarised in
Table~\ref{fitresults}. The statistical uncertainty is obtained from
the fit  and the systematic uncertainty is evaluated as
described in section~\ref{systematic}. \par
The fitted theory generally describes the data at hadron level well
within the fit ranges, as seen in  Fig.~\ref{fit} and confirmed by the
\chisqd\ values in Table~\ref{fitresults}.  The \chisqd\ values are 
based only on the statistical uncertainties of the data and thus it is reasonable
that they are larger than unity.  The extrapolations outside of the
fit regions also provide a reasonable description of the data.  \par
For comparison the numerical results for fits using NLO+NLLA+K predictions are
compiled in Table~\ref{fitresults_nlo+nlla}. The values of \as\ are consistently 
smaller and the theoretical uncertainty is about a factor of 
three larger compared to the fit using NNLO+NLLA+K calculations.
The fit quality reflected by the \chisqd-value is similar for
both QCD predictions, but slightly better for the fit based on NNLO+NLLA+K predictions. 
The difference between the \as\ value returned 
with matched NLO and matched NNLO predictions is of similar
size as the theoretical uncertainty evaluated for the fit using
NLO+NLLA+K calculations. The reduction of the theoretical 
uncertainty using NNLO+NLLA+K QCD predictions
is associated to the higher-order terms available
in the new calculations. \par
The result obtained here cannot be compared directly to the measurement 
of \as\ using the identical data set and the \yy\ event shape 
observable~\cite{Bethke:2008hf}. While for the \as\ measurement 
in~\cite{Bethke:2008hf} matched NNLO+NLLA QCD predictions 
are used, in this analysis NNLO+NLLA+K predictions are applied, 
which take subleading logarithms into account.
\begin{table}[h]
\begin{center}
\begin{tabular}[tbp]{|c|r|r|r|r|r|r|r|} \hline
\rs\ [GeV] & \asrs & stat. & exp. & hadr. & scale &\chisqd \\
\hline
$14.00$  & $  0.1704$ &  $  0.0029$ & $  0.0019$ &  $  0.0079$ & $  0.0028$ &  $62.95/10$ \\ 
$22.00$  & $  0.1562$ &  $  0.0044$ & $  0.0030$ &  $  0.0128$ & $  0.0013$ &  $13.63/10$ \\ 
$34.60$  & $  0.1399$ &  $  0.0015$ & $  0.0023$ &  $  0.0086$ & $  0.0009$ &  $13.17/10$ \\ 
$35.00$  & $  0.1469$ &  $  0.0013$ & $  0.0043$ &  $  0.0086$ & $  0.0011$ &  $11.69/10$ \\ 
$38.30$  & $  0.1375$ &  $  0.0052$ & $  0.0084$ &  $  0.0096$ & $  0.0013$ &  $37.58/10$ \\ 
$43.80$  & $  0.1329$ &  $  0.0029$ & $  0.0032$ &  $  0.0045$ & $  0.0008$ &  $20.29/10$ \\ 

\hline
\end{tabular}
\end{center}
\caption{The value of \as\ using matched NNLO+NLLA+K predictions together with the statistical,
experimental, hadronisation and theoretical uncertainties as described
in section ~\ref{systematic}. 
The last column shows the \chisqd\ value of the fit obtained with statistical uncertainties only 
at the respective energy points.}
\label{fitresults}
\end{table}
\begin{table}[h]
\begin{center}
\begin{tabular}[tbp]{|c|r|r|r|r|r|r|r|r|} \hline
\rs\ [GeV] & \asrs & stat. & exp. & hadr.& scale  &\chisqd \\
\hline
$14.00$  & $  0.1648$ &  $  0.0026$ & $  0.0020$ &  $  0.0068$ & $  0.0039$ &  $64.44/10$ \\ 
$22.00$  & $  0.1518$ &  $  0.0040$ & $  0.0026$ &  $  0.0118$ & $  0.0043$ &  $13.12/10$ \\ 
$34.60$  & $  0.1367$ &  $  0.0014$ & $  0.0022$ &  $  0.0082$ & $  0.0025$ &  $12.39/10$ \\ 
$35.00$  & $  0.1432$ &  $  0.0012$ & $  0.0039$ &  $  0.0083$ & $  0.0028$ &  $14.54/10$ \\ 
$38.30$  & $  0.1338$ &  $  0.0048$ & $  0.0075$ &  $  0.0089$ & $  0.0042$ &  $38.47/10$ \\ 
$43.80$  & $  0.1300$ &  $  0.0027$ & $  0.0029$ &  $  0.0043$ & $  0.0021$ &  $21.65/10$ \\ 

\hline
\end{tabular}
\end{center}
\caption{The value of \as\ using NLO+NLLA+K predictions together with the statistical,
experimental, hadronisation and theoretical uncertainty as described
in section~\ref{systematic} . 
The last column shows the \chisqd\ value of the fit obtained with statistical uncertainties only 
at the respective energy points.}
\label{fitresults_nlo+nlla}
\end{table}
\subsection{Combination of \boldmath{\as} Measurements}
\label{combination}
The results of the measurements of the strong coupling \as\ at the
various centre-of-mass energies are combined to a single value of \asmz. 
For this the values of \asrs\ obtained are evolved to a common energy scale
\mz\ and combined using a weighted mean. 
The theoretical uncertainty as well as the uncertainty
originating from modelling the hadronisation process are likewise determined 
by calculating the weighted mean of the uncertainty evaluated at each single 
energy point. The difficulties
arise in estimating the correlations between the 
systematic uncertainties obtained at the different energy points.
The identical problem is present for the combination
of \as\ results from the LEP-collaborations and
for this reason we use the same method as 
outlined in~\cite{Bethke:2008hf,Schieck:2006tc,Abbiendi:2004qz}. For this analysis we combine the
values obtained at centre-of-mass energies between 22 and 43.8~GeV,
excluding the result obtained at 14~GeV because of the large 
hadronisation corrections and the large \chisqd\ returned by the fit. 
The combination of the results obtained between 22 and 43.8~GeV results in \par
\begin{figure}[htb!]
\begin{center}
\includegraphics[width=1.0\textwidth]{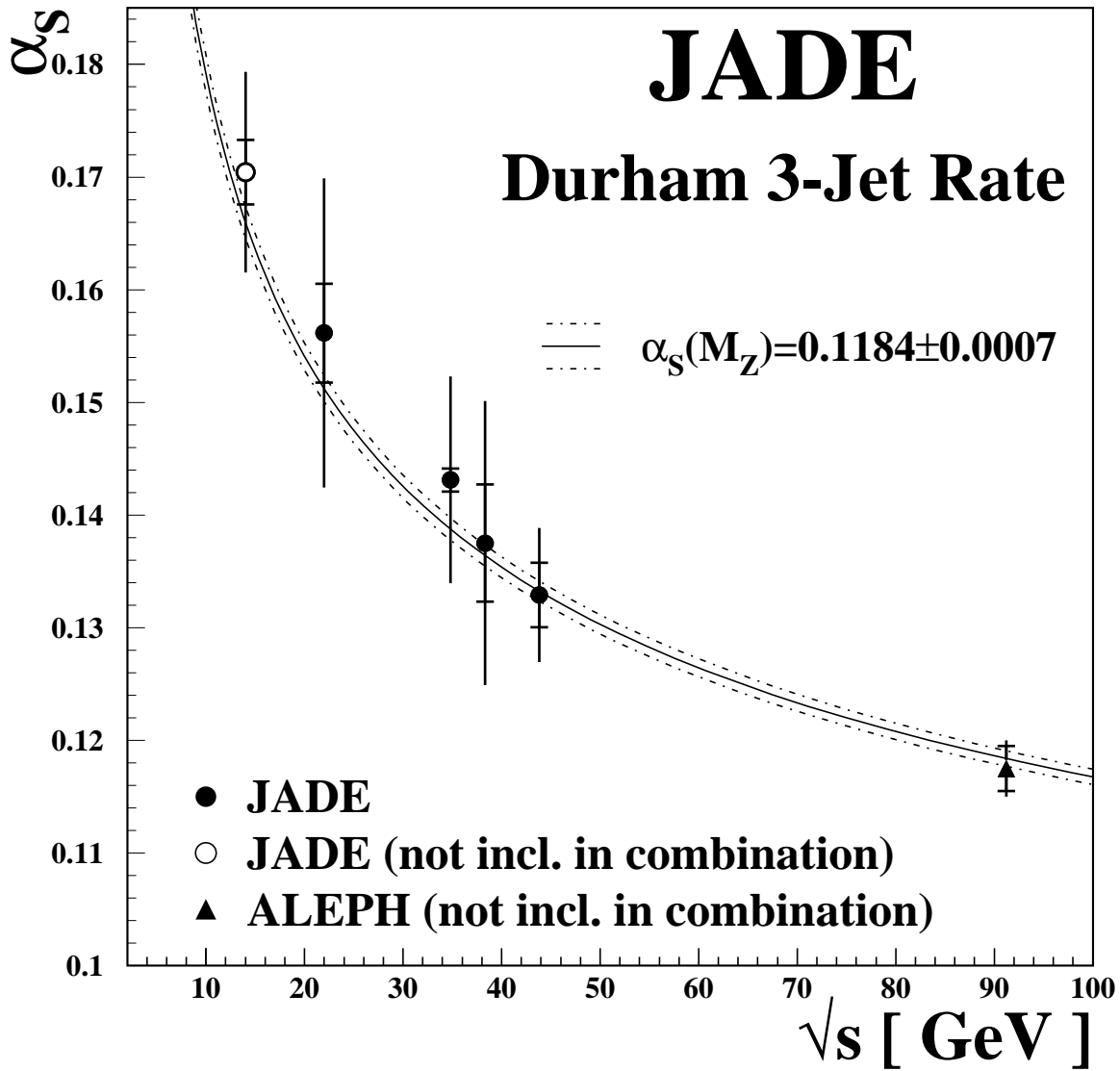}
\end{center}
\caption{
The open and solid points show the measurements of \as\ from the three-jet rate at the various
centre-of-mass energies. The error bars indicate the statistical (inner part) and the
total error. The lines show the world average
of \asmz~\cite{Bethke:2009jm}. The results at 34.6 and 35.0~GeV are
combined to a single value. The result obtained at 14~GeV (open point) is not used 
as input for the  combined value of \as.
The result shown at  a centre-of-mass energy of 91.2~GeV  (triangle) is obtained from 
a fit to the three-jet rate using NNLO predictions only
with data taken by the ALEPH detector~\cite{Dissertori:2009qa}.}
\label{alphas_fit}
\end{figure}

 \result \par
consistent with the world average of $\as=0.1184\pm0.0007$ ~\cite{Bethke:2009jm}. 
To the combined result the value of \as\ measured at 22~GeV contributes with
a weight of 0.13, at 34.6~GeV with 0.23, 
at 35.0~GeV with 0.16, at 38~GeV with 0.07 and at 43.8~GeV with a weight of 0.41.
The results at the various centre-of-mass energies are visualised in 
Fig.~\ref{alphas_fit}. Here, the values obtained at 34.6 and 35.0~GeV are
combined to a single value using the same method as described above \par \resultthirty.
The combination of the \as\ measurements at centre-of-mass energies 
between 22 and 43.8~GeV determined with NLO+NLLA+K predictions leads to \par
\resultnlonlla.
\subsection{Simultaneous Variation of \boldmath{\as} and the Renormalisation Scale }
Besides fixing the renormalisation scale parameter to the natural choice \xmu=1
we repeat the fit as a cross-check with 
both, \as\ and \xmu, being varied within the minimisation procedure, thus choosing the
so called optimal renormalisation scheme~\cite{Bethke:1989jr}. The results of the
fits are summarised in Table~\ref{fitresults_xmuopt}. The
\chisq-values obtained are almost identical to
that from the default fit. The results for the  
scale \xmu=\xmuopt\ at the minimal \chisqd-value vary between $0.31$ and $2.40$, all being 
within the errors consistent with the  \xmu-range used in the systematic
variation of the default fit. To estimate the theoretical uncertainty
the fit is repeated with \xmu\ being set to twice and half of the
value obtained for the optimal scale \xmuopt.
The \as-results at the various energy points between 22 and 43.8~GeV are
combined to a single value using the method described in~\ref{combination}. The
combined value is:\par
\begin{figure}[htb!]
\begin{center}
\begin{tabular}[tbp]{cc}
\includegraphics[width=0.38\textwidth]{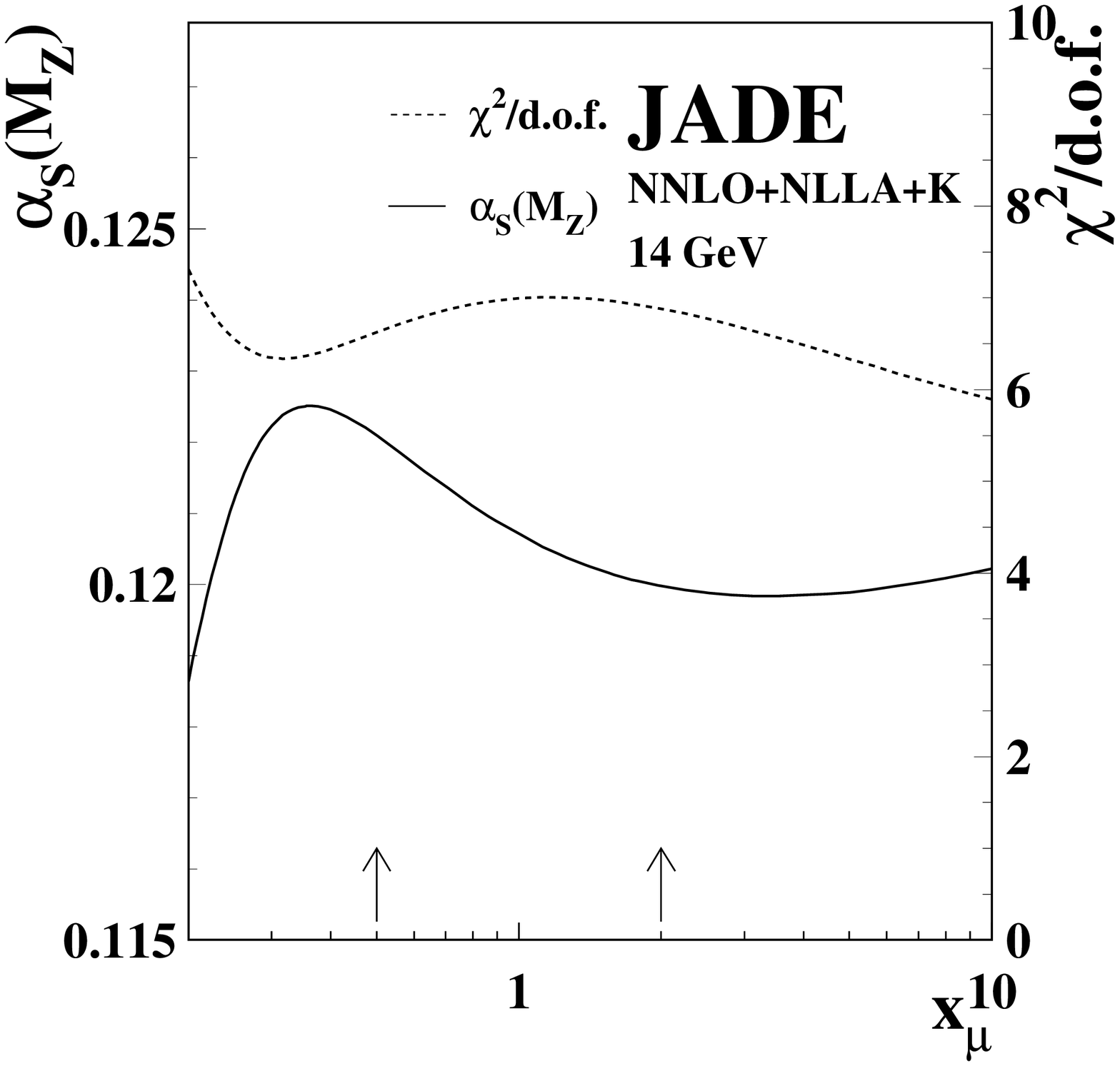} &
\includegraphics[width=0.38\textwidth]{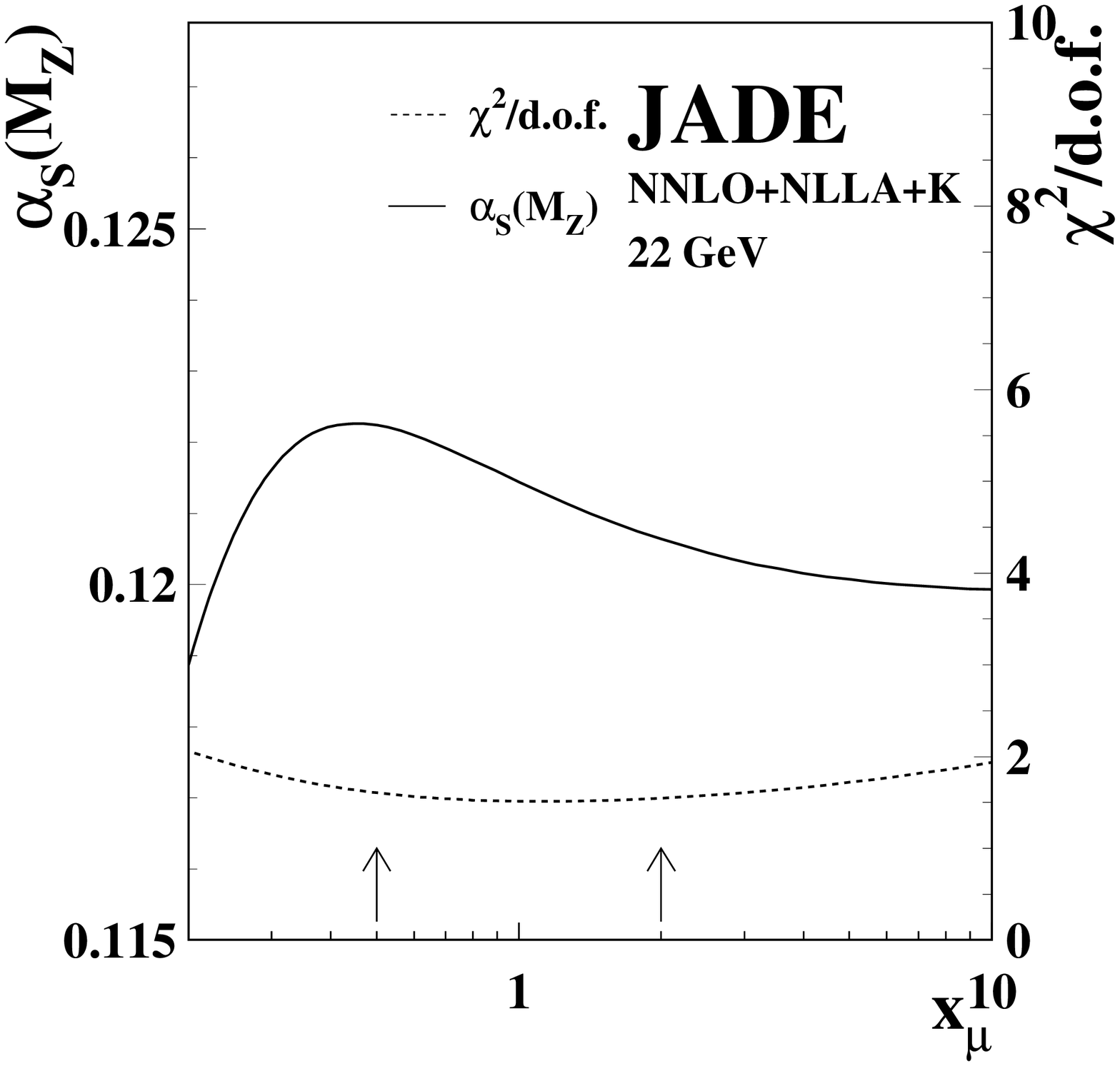}  \\
\includegraphics[width=0.38\textwidth]{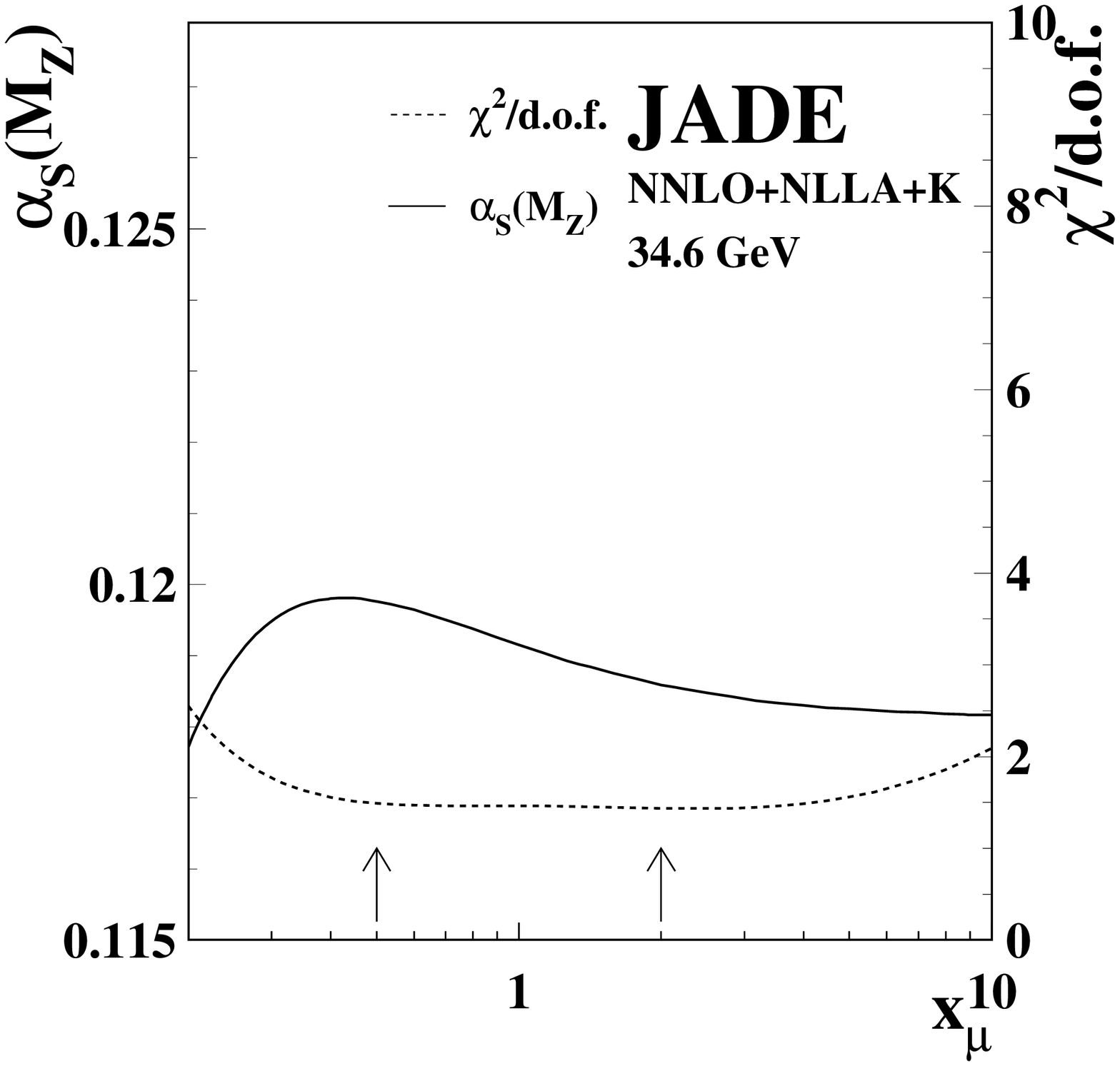} &
\includegraphics[width=0.38\textwidth]{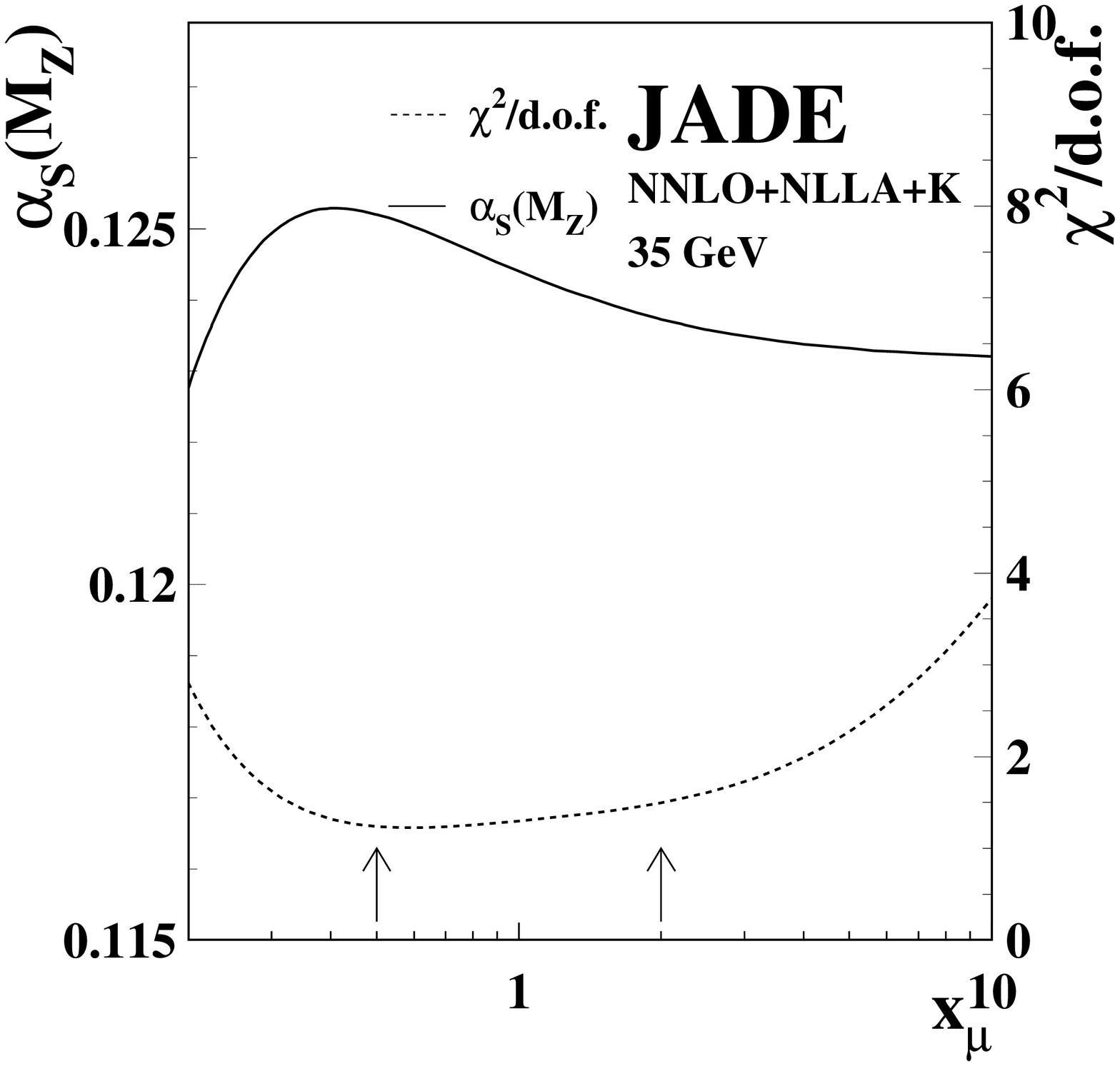}  \\
\includegraphics[width=0.38\textwidth]{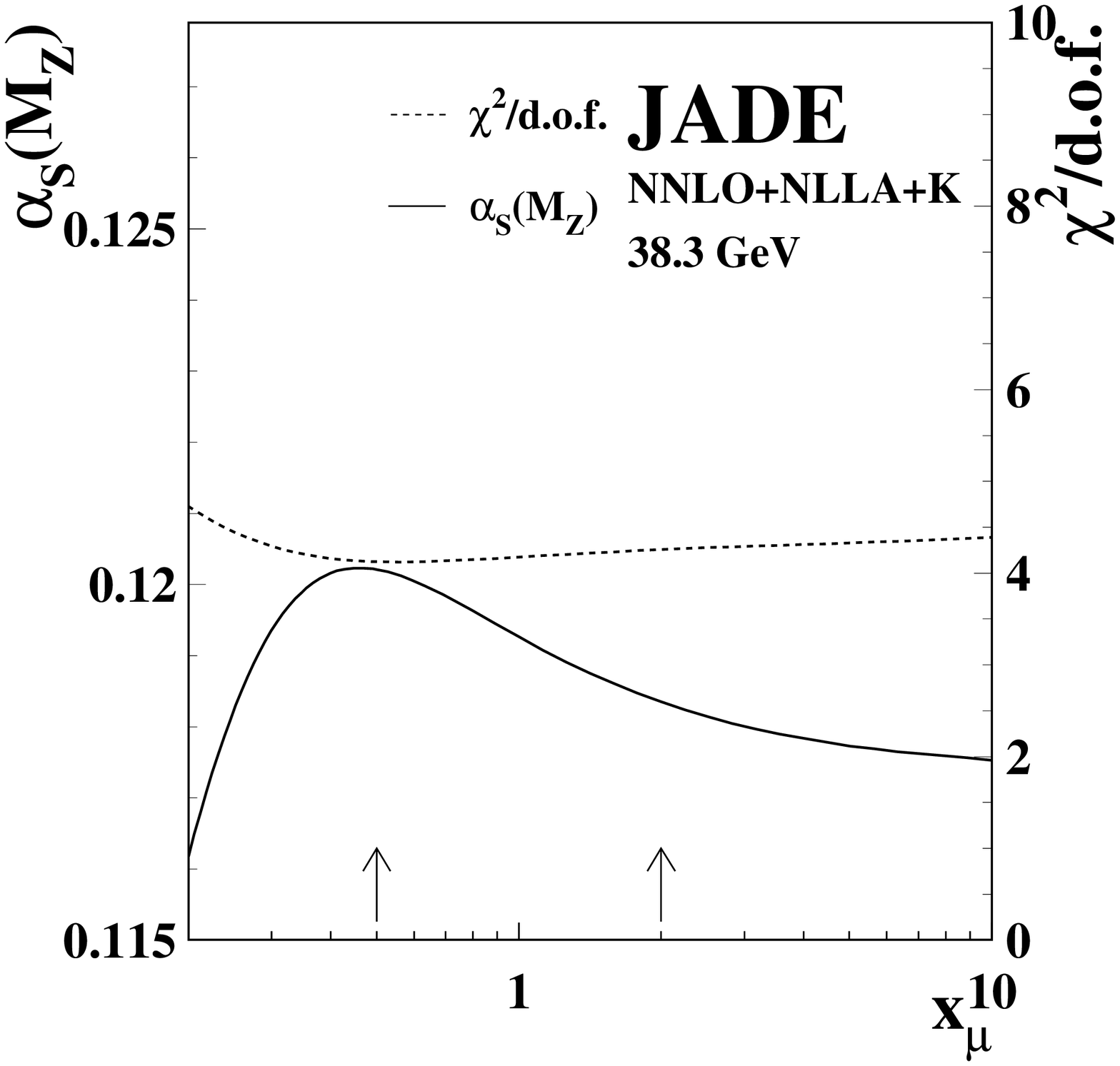} &
\includegraphics[width=0.38\textwidth]{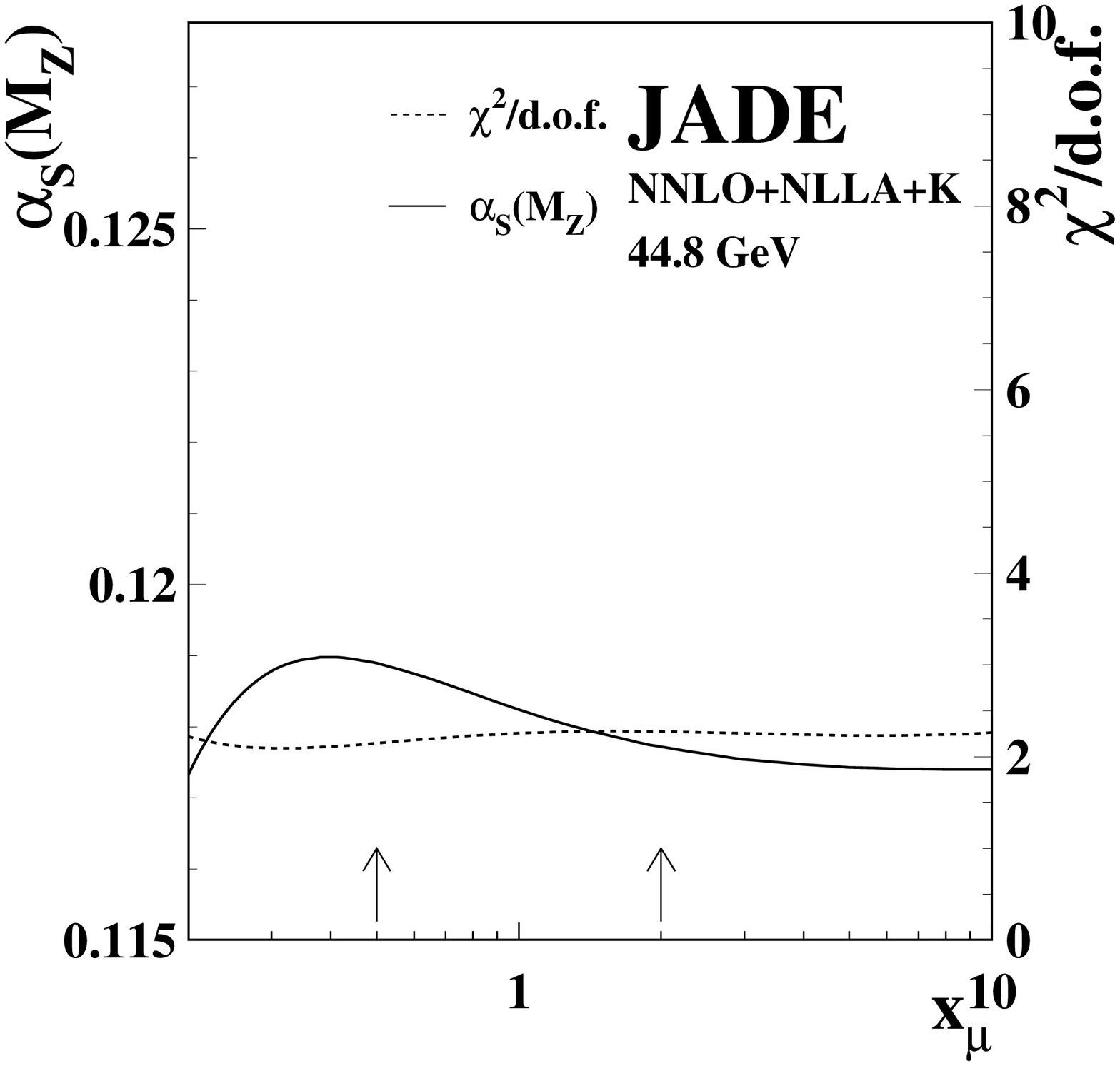}  \\
\end{tabular}
\caption{The variation of the strong coupling \as\ as
a function of the renormalisation scale parameter \xmu\ and the corresponding 
\chisqd-value for the data taken at 
a centre-of-mass energies from 14 to 43.8~GeV using the matched NNLO+NLLA+K predictions. 
The arrows indicate the variation of the renormalisation scale parameter
used to evaluate the theoretical systematic uncertainty.}
\label{asvsxmu}
\end{center}
\end{figure}

\resultxmuopt, \par
leading to almost the identical result as obtained with the default fit.
The variation of \as\ and \chisqd\ with respect to the renormalisation scale is summarised
in Fig.~\ref{asvsxmu}.
The changes for both, the strong
coupling \as\ as well as the \chisqd-value, are small within the
\xmu-range considered, indicating that our results are only moderately sensitive to 
missing higher order terms.
\begin{table}[ht]
\begin{center}
\begin{tabular}[tbp]{|c|r|r|r|r|r|r|r|r|} \hline
\rs\ [GeV] & \asrs & stat. & exp. & hadr.& scale  & \xmuopt &  Corr. &\chisqd \\
\hline
$14.00$  & $  0.1739$ &  $  0.0032$ & $  0.0020$ &  $  0.0088$ & $  0.0149$ &  $ 0.32  \pm 0.06 $ &  0.26 & $57.04/9$ \\ 
$22.00$  & $  0.1559$ &  $  0.0044$ & $  0.0029$ &  $  0.0127$ & $  0.0014$ &  $ 1.11  \pm 1.07 $ &  -0.44 & $13.62/9$ \\ 
$34.60$  & $  0.1390$ &  $  0.0015$ & $  0.0023$ &  $  0.0086$ & $  0.0007$ &  $ 2.40  \pm 2.08 $ &  -0.39 & $12.89/9$ \\ 
$35.00$  & $  0.1478$ &  $  0.0014$ & $  0.0044$ &  $  0.0088$ & $  0.0012$ &  $ 0.58  \pm 0.29 $ &  -0.49 & $11.03/9$ \\ 
$38.30$  & $  0.1387$ &  $  0.0054$ & $  0.0087$ &  $  0.0100$ & $  0.0016$ &  $ 0.55  \pm 0.36 $ &  0.15 & $37.10/9$ \\ 
$43.80$  & $  0.1337$ &  $  0.0029$ & $  0.0034$ &  $  0.0048$ & $  0.0042$ &  $ 0.31  \pm 0.16 $ &  0.27 & $18.82/9$ \\ 

\hline
\end{tabular}
\end{center}
\caption{The value of \as\ using matched NNLO+NLLA+K predictions applying the optimised 
renormalisation scheme. The result is shown together with  the statistical,
experimental, hadronisation, theoretical uncertainties, the 
renormalisation scale parameter \xmuopt, the correlation between \as\ and \xmuopt\  as well as
the \chisqd\ value of the fit.}
\label{fitresults_xmuopt}
\end{table}
\clearpage
\subsection{Measurements of \boldmath{\as} using NNLO Predictions only}
\begin{figure}[h!]
\begin{center}
\includegraphics[width=0.5\textwidth]{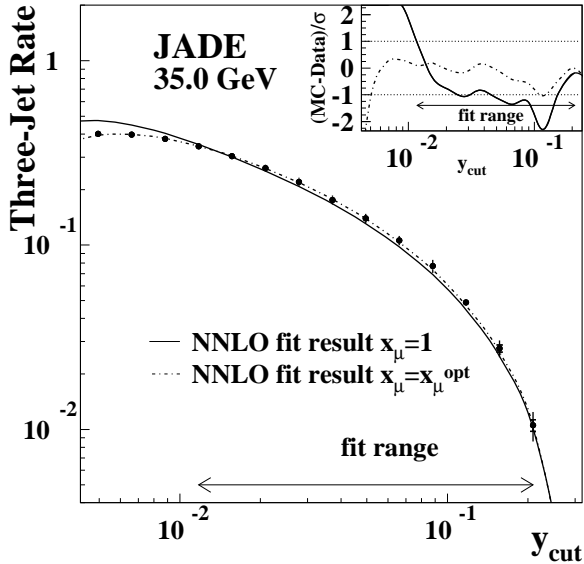} 
\end{center}
\caption{The three-jet rate distribution at a centre-of-mass energy of 35~GeV is shown
together with the NNLO QCD prediction 
 evaluated at the
  \as-value obtained from the least-$\chi^{2}$ fit (solid line).  In addition the 
QCD-predictions  obtained with the optimised
renormalisation scale parameter (dash-dotted line) is shown. The QCD predictions
are corrected for hadronisation effects. 
The insert shows the deviation of the data points from the
QCD-prediction with the \as-value obtained from the fit, normalised
to the combined statistical and experimental error.}
\label{fitNNLO}
\end{figure}

To compare our result with results obtained with NNLO predictions
only, we  repeat the fit without NLLA matching. The fit range is chosen to
be identical to the fit range used in the default fit. 
We perform two different fits using different choices for the renormalisation scale parameter:
a fixed value of \xmu=1 and the optimised renormalisation scheme,
where \xmu\ is an additional free parameter in the fit.
The numbers in Table~\ref{fitresultNNLO} show the result obtained
at a centre-of-mass energy of 35.0~GeV, 
the energy point with the largest number of selected hadronic events. 
The results obtained at all energy points are summarised in the appendix in Table~\ref{fitresults_nnlo} and
Table~\ref{fitresults_nnlo_xmuopt}.
Fig.~\ref{fitNNLO} shows the fit result  using NNLO predictions with \xmu=1 and \xmu=\xmuopt\ at a centre-of-mass 
energy of 35.0~GeV. \par
The shape of the three-jet distribution is not well matched by using
NNLO predictions only, even at large \ycut\ values. 
The \chisqd-value obtained with \xmu=1 increases significantly compared
to the corresponding NNLO+NLLA+K fit.
The fit using the optimised renormalisation returns reasonable \chisqd-values and
smaller \xmu-values. 
A similar behavior was already observed using NLO predictions only with the renormalisation 
factor being set to \xmu=1 and \xmu=\xmuopt\ ~\cite{Bethke:1989jr,Abreu:2000ck} where variation
of the renormalisation scale factor led to an improved description of the data as well.
For both renormalisation scale schemes the scale uncertainty is considerably increased compared
to the measurement using matched NNLO+NLLA+K predictions.
In addition the difference between the fit using the
natural and the optimised renormalisation scale is increased compared to a comparison using
matched NNLO+NLLA+K predictions. 
A large sensitivity to the choice of the fit range is observed for a fit using NNLO predictions only. Again,
the measurement of \as\ with data taken at a centre-of-mass energy of 14~GeV returns by far the
largest \chisqd.  \par
We conclude, contrary to~\cite{Dissertori:2009qa} which performs a fit to a single \ycut-bin only and therefore being insensitive to the shape of the distribution, 
that resummation affects the fit significantly by decreasing the scale dependence and making
the result of the fit  more reliable.
For this reason we consider the result based only on NNLO predictions as a 
cross-check only. 
\begin{table}[h]
\begin{center}
\begin{tabular}[]{|c|c|r|r|r|r|r|r|} \hline
 &\as(35 GeV) & stat. & exp. & hadr. & scale & \xmu &  \chisqd \\
\hline
\xmu\ = 1  & $  0.1426$ &  $  0.0012$ & $  0.0040$ &  $  0.0089$ & $  0.0015$ &  1.0 & $45.30/10$ \\ 
\xmu\ = \xmuopt   & $  0.1486$ &  $  0.0014$ & $  0.0049$ &  $  0.0089$ & $  0.0050$ &  $ 0.23  \pm 0.03 $ & $11.42/9$ \\ 
\hline
\end{tabular}
\end{center}
\caption{The value of \as\ determined with NNLO predictions only together with the statistical,
experimental, hadronisation and theoretical uncertainty as described
in section ~\ref{systematic}. The last columns show the value for \xmu\ and the \chisqd~The first row shows the fit result for a fixed renormalisation
scale parameter \xmu=1, the second row for \xmu\ being varied within the fit.
The \chisq-value is obtained with the statistical
uncertainty only, taking bin-to-bin correlations into account.}
\label{fitresultNNLO}
\end{table}
\section{Summary}
\begin{figure}[htb!]
\begin{center}
\includegraphics[width=.75\textwidth]{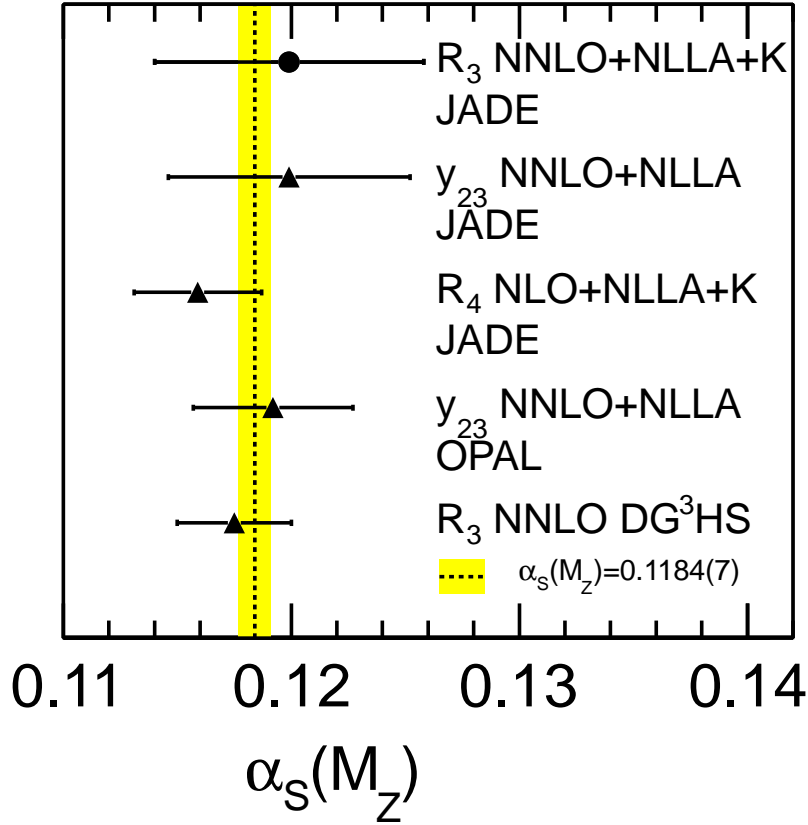}
\end{center}
\caption{Comparison of the \as\ measurement obtained in this analysis with previous \as\ measurements
using the three-jet rate, the four-jet rate and the \yy-differential distribution in \epem-annihilation between 14 and 91~GeV using matched 
and unmatched  QCD predictions. The $1^{\mathrm{st}}$  measurement (top down) is the result of the present analysis, 
the $2^{\mathrm{nd}}$ is the measurement using the differential two-jet rate distribution \yy\ with NNLO+NLLA calculations 
and data taken between 14 and 44~GeV~\cite{Bethke:2008hf}, 
the $3^{\mathrm{rd}}$ is the measurement using the four-jet rate using matched NLO+NLLA+K predictions 
and data between 14 and 44~GeV~\cite{Schieck:2006tc}, 
the $4^{\mathrm{th}}$ is the measurement  using
the differential two-jet rate distribution \yy\ with NNLO+NLLA calculations  and data taken at 91~GeV~\cite{Abbiendi:2011nnlo} and  the 
$5^{\mathrm{th}}$ is a 
measurement using the three-jet rate at a centre-of-mass energy of 91~GeV using NNLO predictions only~\cite{Dissertori:2009qa}. }
\label{SummaryAlphaS}
\end{figure}
In this paper we present a measurement of the strong coupling \as\ using
the three-jet rate  taken
with the JADE experiment at  centre-of-mass energies between 14 and
43.8~GeV. The three-jet rate is compared to simulated events
obtained with \py, \hw\ and \ar. All Monte Carlo
models reproduce the measured distributions well. The strong
coupling \as\ is measured by a fit to matched NNLO+NLLA+K 
predictions. The
combined value using the measurements between 22 and 43.8~GeV results
in \restot. The value is consistent with the world average~\cite{Bethke:2009jm}.
The theory uncertainty has shrunk considerably by using matched NNLO+NLLA+K 
calculations and among the uncertainties it is the smallest.
The dominant uncertainty originates from applying
different Monte Carlo models to estimate the transition
from partons to hadrons. A fit using NNLO predictions only with the renormalisation scale parameter 
\xmu\ set to one cannot
describe the shape of the three-jet rate distribution well. The shape can only be
described well if \as\ and \xmu\ are fitted simultaneously. \par
Fig.~\ref{SummaryAlphaS} shows a comparison
of the \as\ measurement presented in this paper with 
previous \as\ measurements using
higher  QCD predictions.
The results obtained with a fit to the \yy-distribution using data taken with the JADE~\cite{Bethke:2008hf} and the OPAL experiments~\cite{Abbiendi:2011nnlo}
are expected to be strongly correlated with the three-jet rate and 
are considered as a good cross check. The measured mean values, similar to the result obtained with this analysis, 
are slightly above the world average value. The measurements using the four-jet rate and data taken 
with the JADE experiment~\cite{Schieck:2006tc} and the three-jet rate using data taken with the ALEPH experiment~\cite{Dissertori:2009qa} are consistent 
within the uncertainties. All measurements are in agreement with the world average value of
\as~\cite{Bethke:2009jm}.
\clearpage
\section*{Acknowledgements} 
We thank the JADE collaboration for setting up the
terms and for allowing the use of the original JADE data and
software for actual physics analyses. We are especially grateful
to Jan Olsson and to Pedro Movilla Fernandez for their irreplaceable
and invaluable 
efforts and contributions to save and recuperate the data, as well as the
software necessary to analyse these data on modern computer systems. \par
This research was supported by the DFG cluster of excellence "Origin and Structure of the Universe" (www.universe-cluster.de) and by the T\'AMOP 4.2.1./B-09/1/KONV-2010-0007 project and
the Hungarian Scientific Research Fund grant K-101482. \par

\appendix
\section{Tables with hadron level values}
\begin{table}[]
 \begin{center}
\begin{tabular}[tbp]{|c|c|c|c|} \hline
 $\log_{10}(\ycut)$ & $R_{3}$(14~GeV) & $R_{3}$(22~GeV) & $R_{3}$(34.6~GeV) \\
\hline
 $ -4.93 $ &&  &  \\ 
 $ -4.80 $ &&  &  \\ 
 $ -4.68 $ &&  &  \\ 
 $ -4.55 $ &&  &  \\ 
 $ -4.43 $ &&  & $ 0.002  \pm 0.000 \pm 0.003 $  \\ 
$ -4.30 $ &&  & $ 0.001  \pm 0.000 \pm 0.001 $  \\ 
$ -4.18 $ &&  & $ 0.001  \pm 0.000 \pm 0.001 $  \\ 
$ -4.05 $ && $ 0.002  \pm 0.001 \pm 0.002 $ & $ 0.002  \pm 0.000 \pm 0.003 $  \\ 
$ -3.93 $ && $ 0.001  \pm 0.001 \pm 0.001 $ & $ 0.002  \pm 0.000 \pm 0.001 $  \\ 
$ -3.81 $ && $ 0.001  \pm 0.001 \pm 0.001 $ & $ 0.002  \pm 0.000 \pm 0.001 $  \\ 
$ -3.68 $ && $ 0.000  \pm 0.000 \pm 0.004 $ & $ 0.002  \pm 0.000 \pm 0.001 $  \\ 
$ -3.56 $ & $ 0.001  \pm 0.001 \pm 0.001 $ & $ 0.000  \pm 0.001 \pm 0.000 $ & $ 0.003  \pm 0.001 \pm 0.001 $  \\ 
$ -3.43 $ & $ 0.002  \pm 0.001 \pm 0.002 $ & $ 0.001  \pm 0.001 \pm 0.001 $ & $ 0.006  \pm 0.001 \pm 0.002 $  \\ 
$ -3.31 $ & $ 0.002  \pm 0.001 \pm 0.001 $ & $ 0.003  \pm 0.002 \pm 0.002 $ & $ 0.014  \pm 0.001 \pm 0.002 $  \\ 
$ -3.18 $ & $ 0.003  \pm 0.001 \pm 0.001 $ & $ 0.006  \pm 0.002 \pm 0.002 $ & $ 0.026  \pm 0.001 \pm 0.002 $  \\ 
$ -3.06 $ & $ 0.005  \pm 0.002 \pm 0.001 $ & $ 0.013  \pm 0.003 \pm 0.002 $ & $ 0.053  \pm 0.002 \pm 0.005 $  \\ 
$ -2.93 $ & $ 0.008  \pm 0.002 \pm 0.001 $ & $ 0.025  \pm 0.004 \pm 0.003 $ & $ 0.098  \pm 0.003 \pm 0.006 $  \\ 
$ -2.81 $ & $ 0.015  \pm 0.003 \pm 0.003 $ & $ 0.047  \pm 0.006 \pm 0.003 $ & $ 0.167  \pm 0.003 \pm 0.007 $  \\ 
$ -2.68 $ & $ 0.027  \pm 0.004 \pm 0.007 $ & $ 0.103  \pm 0.008 \pm 0.009 $ & $ 0.251  \pm 0.004 \pm 0.011 $  \\ 
$ -2.56 $ & $ 0.061  \pm 0.005 \pm 0.009 $ & $ 0.168  \pm 0.010 \pm 0.013 $ & $ 0.331  \pm 0.004 \pm 0.012 $  \\ 
$ -2.43 $ & $ 0.109  \pm 0.007 \pm 0.014 $ & $ 0.252  \pm 0.012 \pm 0.018 $ & $ 0.393  \pm 0.004 \pm 0.013 $  \\ 
$ -2.31 $ & $ 0.188  \pm 0.009 \pm 0.014 $ & $ 0.343  \pm 0.013 \pm 0.026 $ & $ 0.414  \pm 0.004 \pm 0.009 $  \\ 
$ -2.18 $ & $ 0.281  \pm 0.010 \pm 0.009 $ & $ 0.424  \pm 0.014 \pm 0.015 $ & $ 0.401  \pm 0.004 \pm 0.005 $  \\ 
$ -2.06 $ & $ 0.374  \pm 0.011 \pm 0.019 $ & $ 0.436  \pm 0.014 \pm 0.022 $ & $ 0.369  \pm 0.004 \pm 0.005 $  \\ 
$ -1.93 $ & $ 0.461  \pm 0.011 \pm 0.022 $ & $ 0.413  \pm 0.014 \pm 0.010 $ & $ 0.329  \pm 0.004 \pm 0.009 $  \\ 
$ -1.80 $ & $ 0.485  \pm 0.011 \pm 0.015 $ & $ 0.343  \pm 0.013 \pm 0.017 $ & $ 0.286  \pm 0.004 \pm 0.005 $  \\ 
$ -1.68 $ & $ 0.448  \pm 0.011 \pm 0.007 $ & $ 0.303  \pm 0.013 \pm 0.015 $ & $ 0.248  \pm 0.004 \pm 0.007 $  \\ 
$ -1.55 $ & $ 0.360  \pm 0.011 \pm 0.010 $ & $ 0.240  \pm 0.012 \pm 0.010 $ & $ 0.207  \pm 0.004 \pm 0.008 $  \\ 
$ -1.43 $ & $ 0.281  \pm 0.010 \pm 0.011 $ & $ 0.190  \pm 0.011 \pm 0.010 $ & $ 0.171  \pm 0.003 \pm 0.006 $  \\ 
$ -1.30 $ & $ 0.205  \pm 0.009 \pm 0.013 $ & $ 0.154  \pm 0.010 \pm 0.012 $ & $ 0.132  \pm 0.003 \pm 0.006 $  \\ 
$ -1.18 $ & $ 0.132  \pm 0.008 \pm 0.016 $ & $ 0.115  \pm 0.009 \pm 0.008 $ & $ 0.100  \pm 0.003 \pm 0.005 $  \\ 
$ -1.05 $ & $ 0.084  \pm 0.006 \pm 0.012 $ & $ 0.078  \pm 0.007 \pm 0.006 $ & $ 0.068  \pm 0.002 \pm 0.003 $  \\ 
$ -0.93 $ & $ 0.045  \pm 0.005 \pm 0.011 $ & $ 0.052  \pm 0.006 \pm 0.008 $ & $ 0.044  \pm 0.002 \pm 0.001 $  \\ 
$ -0.81 $ & $ 0.012  \pm 0.002 \pm 0.007 $ & $ 0.030  \pm 0.005 \pm 0.006 $ & $ 0.024  \pm 0.001 \pm 0.002 $  \\ 
$ -0.68 $ & $ 0.005  \pm 0.002 \pm 0.002 $ & $ 0.014  \pm 0.003 \pm 0.002 $ & $ 0.009  \pm 0.001 \pm 0.001 $  \\ 
$ -0.56 $ & $ 0.003  \pm 0.001 \pm 0.002 $ &  &   $ 0.001  \pm 0.000 \pm 0.000 $ \\ 

 \hline
 \end{tabular}
 \end{center}
 \caption{
 Hadron-level value of the three-jet fraction
 using the Durham algorithm at 14, 22 and 34.6 GeV. 
 The value is corrected for contributions from $\epem \to \bbbar$-events.
 In all cases the first quoted error indicates the statistical
 error while the second quoted error corresponds to the
 total experimental uncertainty. Uncertainties consistent with zero
 indicate that the corresponding value is smaller than
 the precision shown in the Table.}
 \label{hadron_tab_1}
 \end{table}

\clearpage

\begin{table}[]
 \begin{center}
 \begin{tabular}[tbp]{|c|c|c|c|} \hline
 $\log_{10}(\ycut)$ & $R_{3}$(35~GeV) & $R_{3}$(38.3~GeV) & $R_{3}$(43.8~GeV) \\
 \hline
 $ -4.93 $ &&  &  \\ 
 $ -4.80 $ &&  &  \\ 
 $ -4.68 $ &&  &  \\ 
 $ -4.55 $ &&  &  \\ 
 $ -4.43 $ &&  &  \\ 
 $ -4.30 $ &&  &  \\ 
 $ -4.18 $ &&  & $ 0.003  \pm 0.001 \pm 0.006 $  \\ 
$ -4.05 $ & &  & $ 0.005  \pm 0.001 \pm 0.006 $  \\ 
$ -3.93 $ & $ 0.001  \pm 0.000 \pm 0.001 $ & $ 0.001  \pm 0.001 \pm 0.000 $ & $ 0.003  \pm 0.001 \pm 0.003 $  \\ 
$ -3.81 $ & $ 0.001  \pm 0.000 \pm 0.001 $ & $ 0.001  \pm 0.001 \pm 0.001 $ & $ 0.003  \pm 0.001 \pm 0.001 $  \\ 
$ -3.68 $ & $ 0.002  \pm 0.000 \pm 0.001 $ & $ 0.003  \pm 0.002 \pm 0.002 $ & $ 0.005  \pm 0.001 \pm 0.001 $  \\ 
$ -3.56 $ & $ 0.003  \pm 0.000 \pm 0.001 $ & $ 0.005  \pm 0.002 \pm 0.002 $ & $ 0.007  \pm 0.001 \pm 0.001 $  \\ 
$ -3.43 $ & $ 0.006  \pm 0.001 \pm 0.001 $ & $ 0.008  \pm 0.003 \pm 0.003 $ & $ 0.012  \pm 0.002 \pm 0.002 $  \\ 
$ -3.31 $ & $ 0.011  \pm 0.001 \pm 0.001 $ & $ 0.018  \pm 0.004 \pm 0.003 $ & $ 0.026  \pm 0.003 \pm 0.003 $  \\ 
$ -3.18 $ & $ 0.024  \pm 0.001 \pm 0.001 $ & $ 0.040  \pm 0.005 \pm 0.007 $ & $ 0.053  \pm 0.004 \pm 0.004 $  \\ 
$ -3.06 $ & $ 0.050  \pm 0.002 \pm 0.003 $ & $ 0.064  \pm 0.007 \pm 0.005 $ & $ 0.101  \pm 0.005 \pm 0.007 $  \\ 
$ -2.93 $ & $ 0.096  \pm 0.002 \pm 0.002 $ & $ 0.123  \pm 0.009 \pm 0.009 $ & $ 0.172  \pm 0.006 \pm 0.007 $  \\ 
$ -2.81 $ & $ 0.165  \pm 0.003 \pm 0.004 $ & $ 0.187  \pm 0.011 \pm 0.010 $ & $ 0.254  \pm 0.007 \pm 0.011 $  \\ 
$ -2.68 $ & $ 0.245  \pm 0.003 \pm 0.008 $ & $ 0.270  \pm 0.012 \pm 0.014 $ & $ 0.339  \pm 0.008 \pm 0.006 $  \\ 
$ -2.56 $ & $ 0.326  \pm 0.004 \pm 0.009 $ & $ 0.346  \pm 0.013 \pm 0.009 $ & $ 0.390  \pm 0.008 \pm 0.008 $  \\ 
$ -2.43 $ & $ 0.381  \pm 0.004 \pm 0.003 $ & $ 0.379  \pm 0.013 \pm 0.006 $ & $ 0.415  \pm 0.008 \pm 0.009 $  \\ 
$ -2.31 $ & $ 0.401  \pm 0.004 \pm 0.004 $ & $ 0.386  \pm 0.013 \pm 0.012 $ & $ 0.411  \pm 0.008 \pm 0.007 $  \\ 
$ -2.18 $ & $ 0.397  \pm 0.004 \pm 0.004 $ & $ 0.365  \pm 0.013 \pm 0.010 $ & $ 0.381  \pm 0.008 \pm 0.007 $  \\ 
$ -2.06 $ & $ 0.377  \pm 0.004 \pm 0.004 $ & $ 0.328  \pm 0.013 \pm 0.014 $ & $ 0.346  \pm 0.008 \pm 0.009 $  \\ 
$ -1.93 $ & $ 0.343  \pm 0.004 \pm 0.008 $ & $ 0.300  \pm 0.013 \pm 0.013 $ & $ 0.307  \pm 0.008 \pm 0.005 $  \\ 
$ -1.80 $ & $ 0.304  \pm 0.004 \pm 0.007 $ & $ 0.279  \pm 0.012 \pm 0.015 $ & $ 0.281  \pm 0.007 \pm 0.011 $  \\ 
$ -1.68 $ & $ 0.261  \pm 0.003 \pm 0.011 $ & $ 0.268  \pm 0.012 \pm 0.017 $ & $ 0.242  \pm 0.007 \pm 0.006 $  \\ 
$ -1.55 $ & $ 0.220  \pm 0.003 \pm 0.011 $ & $ 0.239  \pm 0.012 \pm 0.021 $ & $ 0.195  \pm 0.006 \pm 0.006 $  \\ 
$ -1.43 $ & $ 0.176  \pm 0.003 \pm 0.009 $ & $ 0.198  \pm 0.011 \pm 0.017 $ & $ 0.156  \pm 0.006 \pm 0.006 $  \\ 
$ -1.30 $ & $ 0.140  \pm 0.003 \pm 0.008 $ & $ 0.152  \pm 0.010 \pm 0.015 $ & $ 0.126  \pm 0.005 \pm 0.007 $  \\ 
$ -1.18 $ & $ 0.106  \pm 0.002 \pm 0.006 $ & $ 0.117  \pm 0.009 \pm 0.012 $ & $ 0.097  \pm 0.005 \pm 0.004 $  \\ 
$ -1.05 $ & $ 0.077  \pm 0.002 \pm 0.006 $ & $ 0.088  \pm 0.008 \pm 0.009 $ & $ 0.072  \pm 0.004 \pm 0.005 $  \\ 
$ -0.93 $ & $ 0.049  \pm 0.002 \pm 0.001 $ & $ 0.060  \pm 0.006 \pm 0.010 $ & $ 0.043  \pm 0.003 \pm 0.004 $  \\ 
$ -0.81 $ & $ 0.028  \pm 0.001 \pm 0.003 $ & $ 0.034  \pm 0.005 \pm 0.007 $ & $ 0.023  \pm 0.003 \pm 0.001 $  \\ 
$ -0.68 $ & $ 0.011  \pm 0.001 \pm 0.002 $ & $ 0.007  \pm 0.002 \pm 0.002 $ & $ 0.008  \pm 0.002 \pm 0.002 $  \\ 
$ -0.56 $ & $ 0.002  \pm 0.000 \pm 0.000 $ & $ 0.006  \pm 0.002 \pm 0.009 $ & $ 0.003  \pm 0.001 \pm 0.002 $  \\ 

 \hline
 \end{tabular}
 \end{center}
 \caption{
 Hadron-level value of the three-jet fraction
 using the Durham algorithm at 35, 38.3 and 43.8 GeV.
 The value is corrected for contributions from $\epem \to \bbbar$-events.
 In all cases the first quoted error indicates the statistical
 error while the second quoted error corresponds to the
 total experimental uncertainty. Uncertainties consistent with zero
 indicate that the corresponding value is smaller than
 the precision shown in the Table.}
 \label{hadron_tab_2}
 \end{table}

\clearpage

\section{Hadronisation correction estimated with simulated events} 

\begin{figure}[!htb]
\begin{center}
\begin{tabular}[tbp]{cc}
\includegraphics[width=0.38\textwidth]{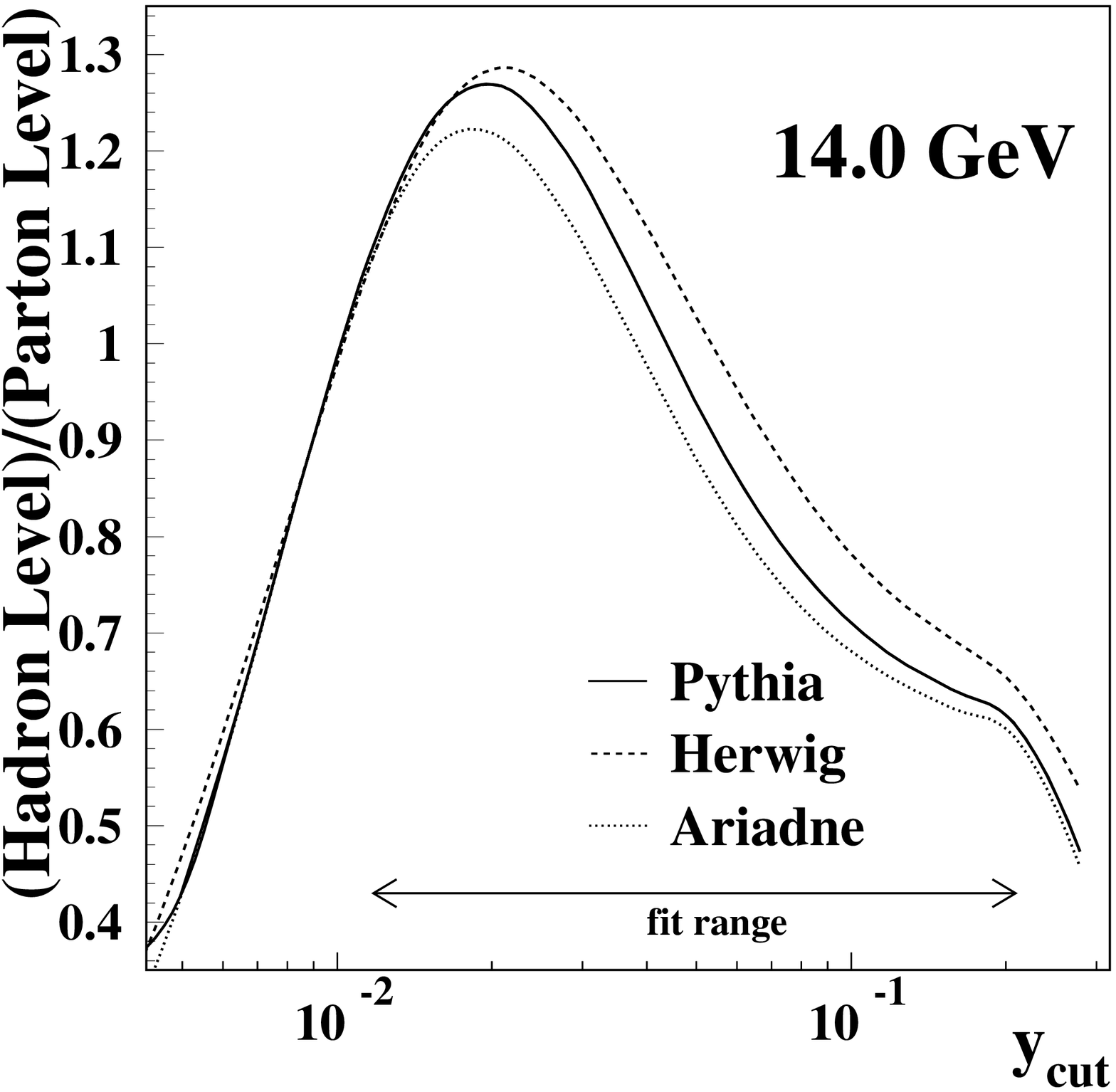} &
\includegraphics[width=0.38\textwidth]{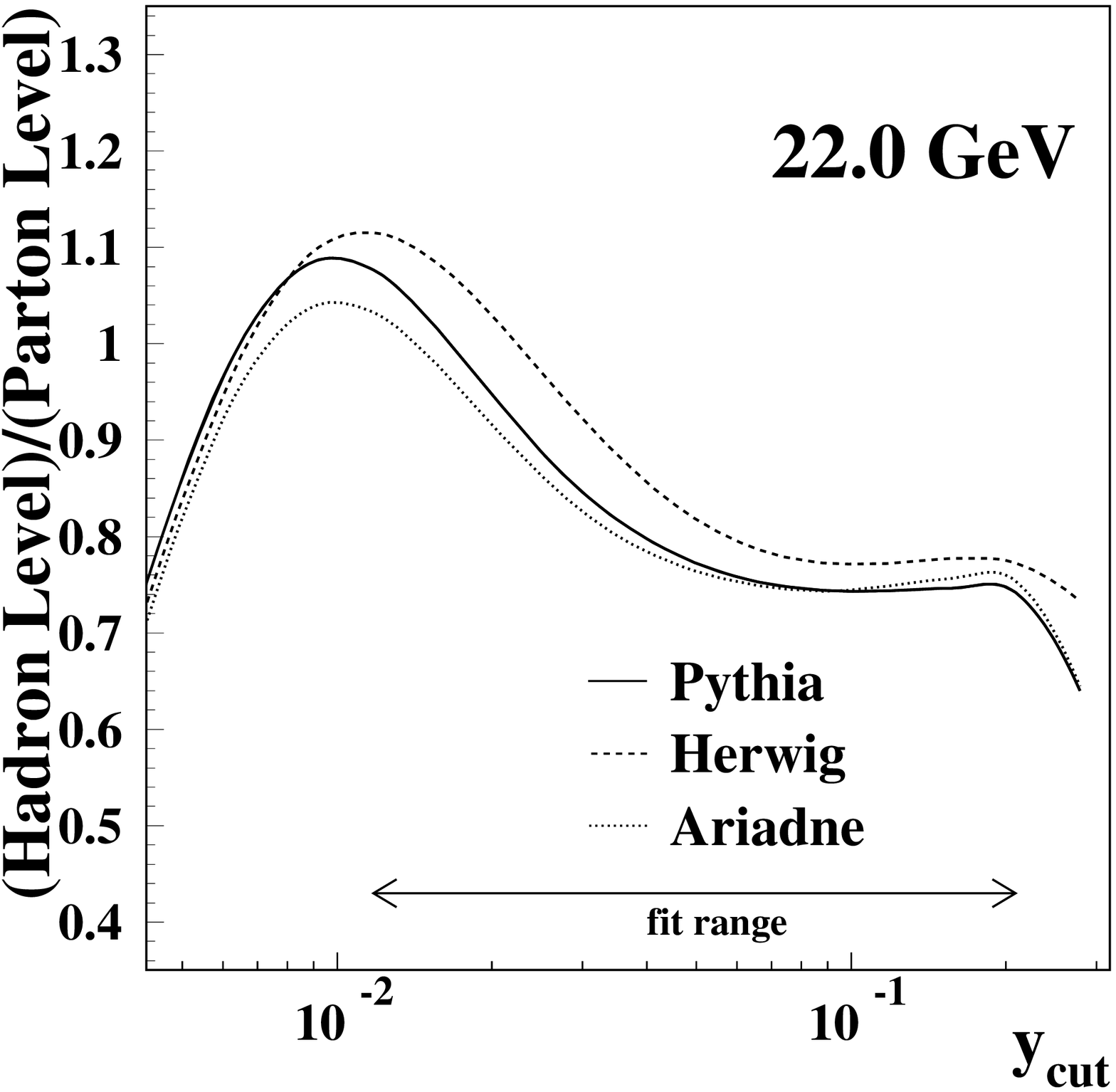}  \\
\includegraphics[width=0.38\textwidth]{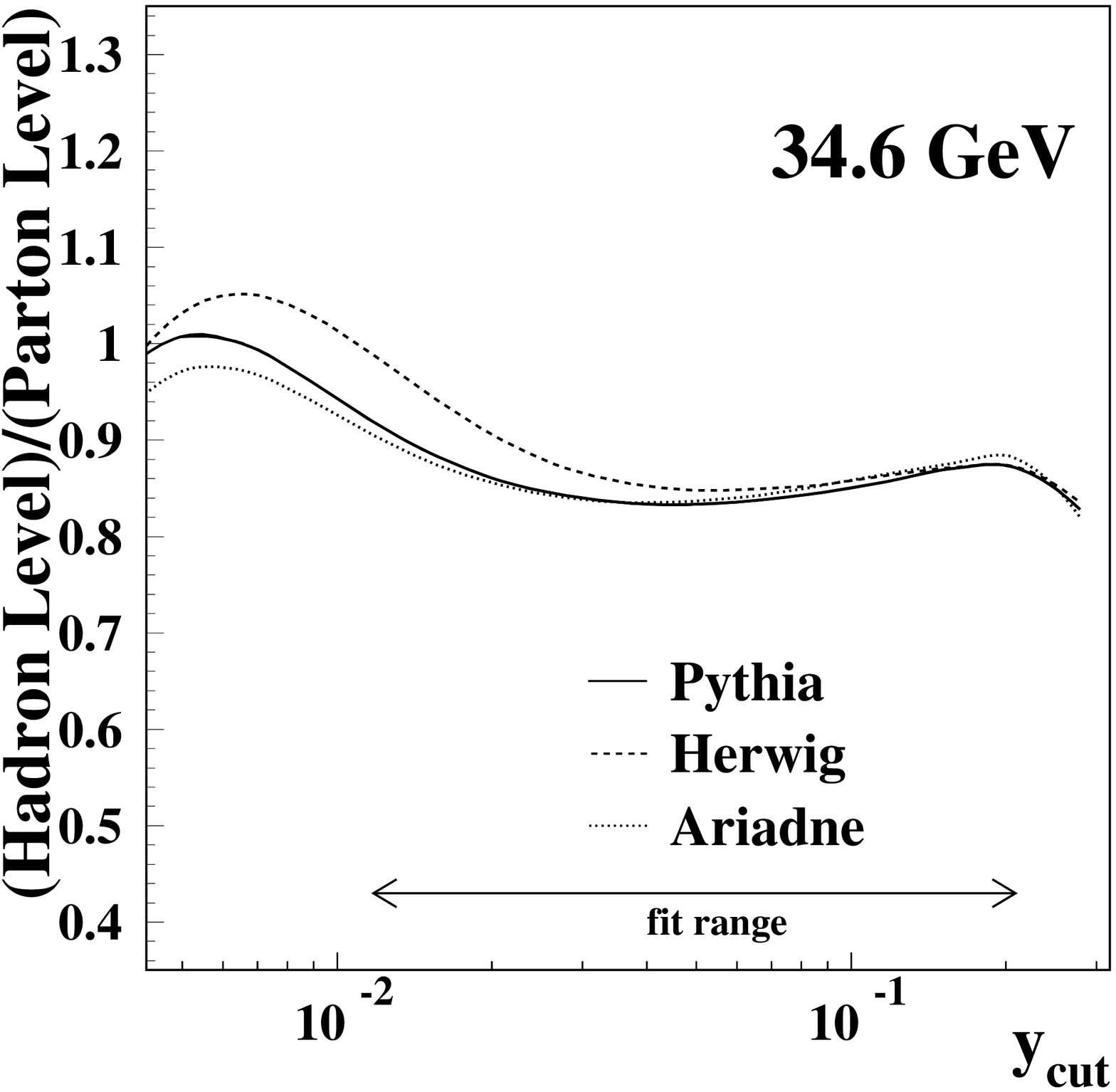} &
\includegraphics[width=0.38\textwidth]{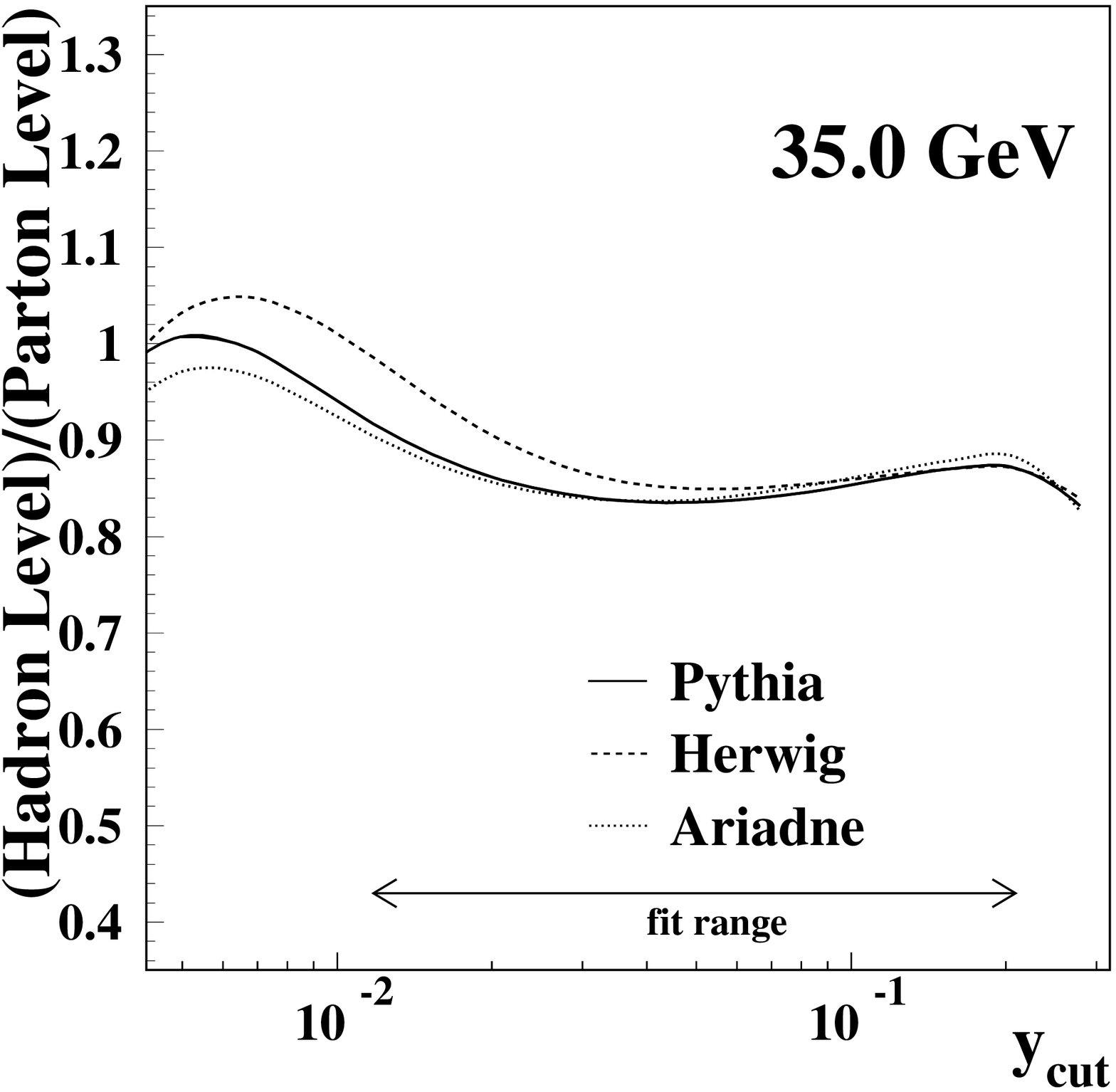}  \\
\includegraphics[width=0.38\textwidth]{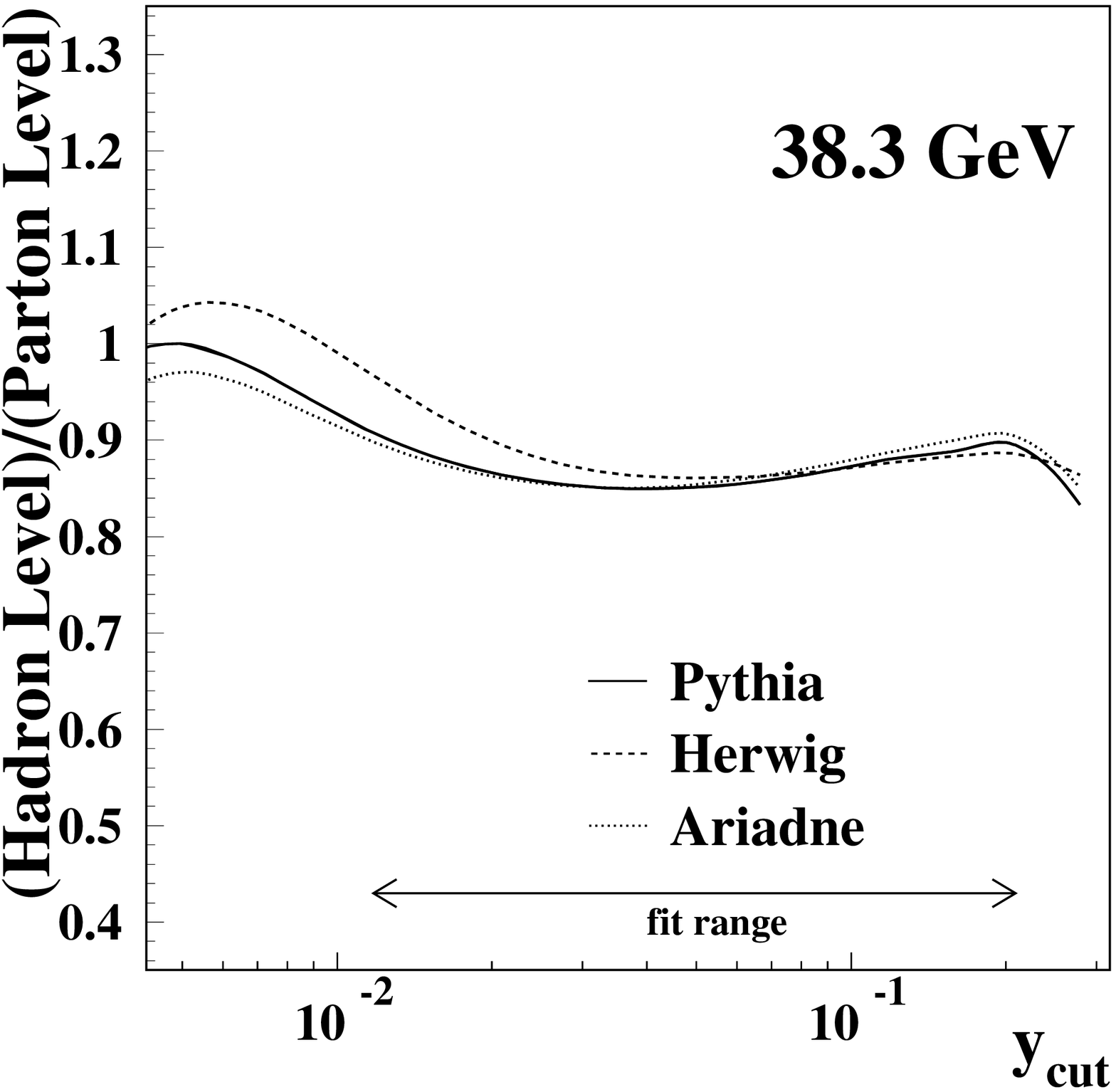} &
\includegraphics[width=0.38\textwidth]{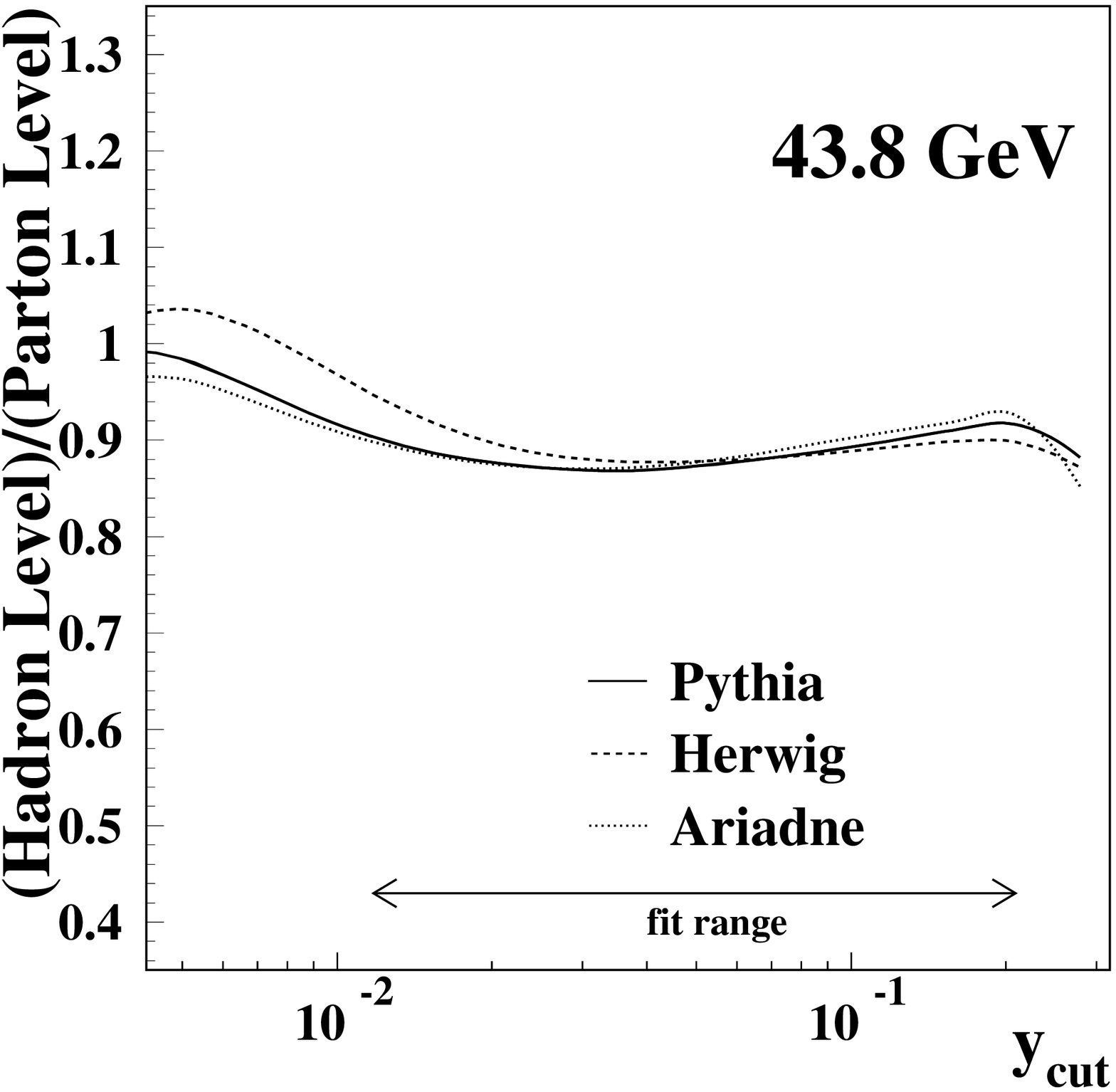} \\
\end{tabular}
\caption{The ratio of the {\em hadron level} prediction divided by the {\em parton level} prediction
as a function of the resolution parameter \ycut\ shown for centre-of-mass energies \rs\ between 14 and 43.8~GeV.}
\label{Hadron_Correction}
\end{center}
\end{figure}



 \clearpage

\section{Fit results with NNLO predictions}

\label{TableNNLO}
\label{appendix_nnlo}
\begin{table}[h]
\begin{center}
\begin{tabular}[tbp]{|c|r|r|r|r|r|r|r|} \hline
\rs\ [GeV] & \asrs & stat. & exp. & hadr. & scale &\chisqd \\
\hline
$14.00$  & $  0.1629$ &  $  0.0023$ & $  0.0036$ &  $  0.0047$ & $  0.0028$ &  $80.24/10$ \\ 
$22.00$  & $  0.1499$ &  $  0.0037$ & $  0.0039$ &  $  0.0105$ & $  0.0018$ &  $18.40/10$ \\ 
$34.60$  & $  0.1364$ &  $  0.0014$ & $  0.0029$ &  $  0.0087$ & $  0.0013$ &  $32.59/10$ \\ 
$35.00$  & $  0.1426$ &  $  0.0012$ & $  0.0040$ &  $  0.0089$ & $  0.0015$ &  $45.30/10$ \\ 
$38.30$  & $  0.1305$ &  $  0.0044$ & $  0.0064$ &  $  0.0085$ & $  0.0015$ &  $47.55/10$ \\ 
$43.80$  & $  0.1301$ &  $  0.0026$ & $  0.0031$ &  $  0.0050$ & $  0.0010$ &  $25.48/10$ \\ 

\hline
\end{tabular}
\end{center}
\caption{The value of \as\ and the statistical,
experimental, hadronisation and theoretical uncertainty as described
in section~\ref{systematic} using NNLO predictions only. 
The last column shows the \chisqd\ value of the fit
for all energy points.}
\label{fitresults_nnlo}
\end{table}

\begin{table}[h]
\begin{center}
\begin{tabular}[tbp]{|c|r|r|r|r|r|r|r|r|} \hline
\rs\ [GeV] & \asrs & stat. & exp. & hadr.& scale  & \xmuopt &  Corr. &\chisqd \\
\hline
$14.00$  & $  0.1748$ &  $  0.0033$ & $  0.0031$ &  $  0.0094$ & $  0.0174$ &  $ 0.22  \pm 0.02 $ &  -0.68 & $61.29/9$ \\ 
$22.00$  & $  0.1547$ &  $  0.0042$ & $  0.0042$ &  $  0.0123$ & $  0.0045$ &  $ 0.28  \pm 0.12 $ &  -0.77 & $16.01/9$ \\ 
$34.60$  & $  0.1407$ &  $  0.0016$ & $  0.0031$ &  $  0.0087$ & $  0.0053$ &  $ 0.23  \pm 0.05 $ &  -0.75 & $16.83/9$ \\ 
$35.00$  & $  0.1486$ &  $  0.0014$ & $  0.0049$ &  $  0.0088$ & $  0.0050$ &  $ 0.23  \pm 0.03 $ &  -0.72 & $11.42/9$ \\ 
$38.30$  & $  0.1651$ &  $  0.0053$ & $  0.0090$ &  $  0.0068$ & $  0.0274$ &  $ 0.12  \pm 0.01 $ &  0.58 & $23.98/9$ \\ 
$43.80$  & $  0.1396$ &  $  0.0037$ & $  0.0050$ &  $  0.0039$ & $  0.0068$ &  $ 0.14  \pm 0.04 $ &  -0.71 & $16.66/9$ \\ 

\hline
\end{tabular}
\end{center}
\caption{The value of \as\ and the statistical,
experimental, hadronisation, renormalisation scale parameter
\xmuopt, the correlation between \as\ and the 
renormalisation scale parameter and the \chisqd\ value of the fit using NNLO predictions.
The fit is performed using the optimised 
renormalisation scheme for all energy points.}
\label{fitresults_nnlo_xmuopt}
\end{table}


\clearpage

\bibliographystyle{prsty}
\bibliography{biblio}      

\end{document}